\documentclass[fleqn,usenatbib]{mnras}

\usepackage{newtxtext,newtxmath}
\usepackage[T1]{fontenc}

\DeclareRobustCommand{\VAN}[3]{#2}
\let\VANthebibliography\thebibliography
\def\thebibliography{\DeclareRobustCommand{\VAN}[3]{##3}\VANthebibliography}

\usepackage{graphicx}	% Including figure files
\usepackage{amsmath}	% Advanced maths commands
\usepackage{mathtools}

\newcommand{\zref}[1][\prescript{0.1}{}]{#1}

\newcommand{\update}[1]{{#1}}
\newcommand{\newupdate}[1]{{#1}}

\usepackage{subfig}

\title[DESI DR2 Galaxy Luminosity Functions]{DESI DR2 Galaxy Luminosity Functions}

\author[S. G. Moore et al.]{Samuel G. Moore,$^{1}$
Shaun Cole,$^{1}$\thanks{E-mail: shaun.cole@durham.ac.uk}
Michael Wilson,$^{1,2}$
Peder Norberg,$^{1,3}$
John Moustakas,$^{4}$ 
J.~Aguilar,$^5$ \and
S.~Ahlen,$^6$
A.~Anand,$^5$
D.~Bianchi,$^{7,8}$
D.~Brooks,$^{9}$
F.~J.~Castander,$^{10,11}$
T.~Claybaugh,$^{5}$ \and
A.~Cuceu,$^{5}$
A.~de la Macorra,$^{12}$
Arjun~Dey,$^{13}$
Biprateep~Dey,$^{14,15}$
S.~Ferraro,$^{4,16}$
A.~Font-Ribera,$^{17}$ \and
J.~E.~Forero-Romero,$^{18,19}$
E.~Gaztanaga,$^{10,11,20}$
S.~Gontcho A Gontcho,$^{5,21}$
G.~Gutierrez,$^{22}$ \and
H.~K.~Herrera-Alcantar,$^{23,24}$
K.~Honscheid,$^{25,26,27}$
M.~Ishak,$^{28}$
R.~Joyce,$^{13}$
S.~Juneau,$^{13}$
R.~Kehoe,$^{29}$
T.~Kisner,$^{5}$ \and
S.~E.~Koposov,$^{30,31}$
A.~Kremin,$^{5}$
O.~Lahav,$^{9}$
C.~Lamman,$^{27}$
M.~Landriau,$^{5}$
L.~Le~Guillou,$^{32}$
M.~E.~Levi,$^{5}$ \and
J.~R.~Lucey,$^3$
M.~Manera,$^{17,33}$
A.~Meisner,$^{13}$
R.~Miquel,$^{17,34}$
S.~Nadathur,$^{20}$
W.~J.~Percival,$^{35,36,37}$ \and
C.~Poppett,$^{5,16,38}$
F.~Prada,$^{39}$ 
A.~J.~Ross,$^{25,27,40}$
G.~Rossi,$^{41}$
E.~Sanchez,$^{42}$
D.~Schlegel,$^{5}$
M.~Schubnell,$^{43,44}$ \and
H.~Seo,$^{45}$ 
J.~Silber,$^5$ 
D.~Sprayberry,$^{13}$
G.~Tarl\'{e},$^{44}$
B.~A.~Weaver,$^{13}$
R.~H.~Wechsler,$^{46,47,48}$
R.~Zhou,$^{5}$
H.~Zou,$^{49}$
\\
\\
\update{Affiliations are listed at the end of the paper}
}

\date{Accepted 2026 April 29. Received 2026 April 20; in original form 2025 October 30}

\pubyear{2025}

\begin{document}
\label{firstpage}
\pagerange{\pageref{firstpage}--\pageref{lastpage}}
\maketitle

\begin{abstract}
\newupdate{We present galaxy luminosity functions (LFs) for the Dark Energy Spectroscopic Instrument (DESI) DR2 Bright Galaxy Survey (BGS) in the $g$, $r$, $z$, and $w1$ bands over $0.002<z<0.6$. Our analysis uses updated k‑corrections and evolutionary corrections, including new polynomial k‑correction fits derived from BGS Year~1 data that supersede earlier GAMA‑based prescriptions. Exploiting the statistical power of DESI, we measure LFs to very faint magnitudes, reaching $^{0.1}M_r-5\log h\sim-10$. Independent measurements from the North and South survey regions show excellent agreement around the LF knee, but the very small statistical uncertainties reveal that simple analytic forms fail to capture the full LF shape. The bright end departs from a pure exponential decline, while the faint end exhibits complex, non‑power‑law behaviour, including a pronounced upturn at $^{0.1}M_r-5\log h\gtrsim-15$, which is stronger for red galaxies than for blue. We show that our LFs are largely complete for galaxies with surface brightness $\mu_{50}<25$, and that an apparent steepening fainter than $-13$ is driven primarily by local overdensity and fragmentation of large galaxies. A systematic North–South offset at the brightest magnitudes is traced to red galaxies and may reflect shallower North photometry underestimating extended early‑type profiles, although this remains inconclusive. We therefore also provide LFs based on model‑Petrosian magnitudes. Redshift‑splitting reveals small but significant residuals, indicating limitations of a simple global evolutionary model. Using the redshift limits of \citet{Loveday2011}, we find excellent agreement with GAMA, with substantially reduced statistical errors. These measurements provide a precise reference for studies of environmental and population‑dependent LFs and for testing galaxy formation models.
}
\end{abstract}


%%%%%%%%%%%%%%%%%%%%%%%%%%%%%%%%%%%%%%%%%%%%%%%%%%

%%%%%%%%%%%%%%%%% BODY OF PAPER %%%%%%%%%%%%%%%%%%

\section{Introduction} \label{Introduction}

\begin{figure*}
    \centering
    \includegraphics[width=\textwidth]{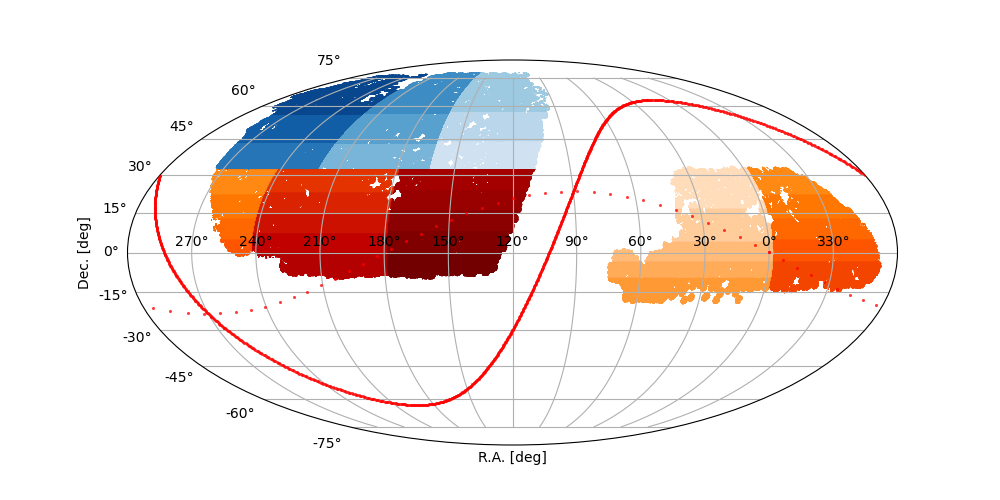}
    \caption{The Y3 DESI Footprint - showing the North (blue) and South (red). The different shades of blue and red highlight the jackknife regions (see Section~\ref{Luminosity Functions} for details). The solid red curve shows the galactic plane while the dotted red curve shows the ecliptic plane.}
    \label{fig:DESI_map}
\end{figure*}

The galaxy luminosity function (LF) has become one of the most useful statistics for quantifying the galaxy population. Various studies have sought to improve the measurement of the LF, including at faint magnitudes. These studies have been facilitated by new generations of redshift surveys which have surveyed large areas of the sky to sufficiently high redshifts with high completeness. This has enabled data analysis on large catalogues of galaxies to produce statistically precise results.

%\todo{ Replace with more comprehensive version. Previous large-scale structure surveys have sought to measure the LF \citep{Cole2001, Norberg2002, 2002AAS...200.2704B, 2011MNRAS.413..971D, Loveday2011} with successive surveys measuring the LF for larger volumes and providing increasing accuracy.  Moreover, as surveys have grown larger, these studies have also probed how the LF varies for different galaxy properties. This has included the LF split by cosmic web classification \citep{10.1093/mnras/stx2638}, field vs. cluster and group classification \citep{10.1093/mnras/stw898}, local density \citep{McNaught-Roberts_2014}, etc. The LF has also been measured in a number of different bands. For example, the Galaxy and Mass Assembly (GAMA) survey has measured the LF in the $ugriz$ bands \citep{Loveday2011}.}

\update{Global LFs have been measured with steadily improving precision as redshift and imaging surveys expanded in area, depth, and wavelength coverage, establishing a broadly consistent empirical picture of the local galaxy population while also revealing persistent tensions at both the bright and faint ends. Early optical determinations such as the CfA \citep{1983ApJS...52...89H} and subsequent local compilations \citep{1992ApJ...390..338L, 1994ApJ...428...43M} were followed by the first “modern” large-sample constraints from surveys such as the Las Campanas Redshift Survey \citep{1996ApJ...470..172S, 1996ApJ...464...60L} and the ESO Slice Project \citep{1997A&A...326..477Z}. By the early 2000s, the very large 2dF Galaxy Redshift Survey and Sloan Digital Sky Survey enabled high signal-to-noise, multi-band LFs with well-characterised selection \citep[e.g:][]{10.1046/j.1365-8711.1999.02721.x, Norberg2002, 2003ApJ...592..819B}, solidifying the empirical conclusion that a Schechter-like form provides an adequate first-order description over a wide luminosity range, while also emphasising that the inferred faint-end slope and normalisation depend sensitively on the treatment of low-luminosity systems. Outside of the optical bands, all-sky and wide-area surveys such as 2MASS have measured near-infrared LFs which more directly reflect galaxy stellar masses  \citep[e.g:][]{2001ApJ...560..566K, Cole2001}.

More recently, global LF work has expanded further from a simple description to focusing on regimes dominated by rare populations. Using SDSS DR6, \cite{2009MNRAS.399.1106M} reported evidence for a statistically significant bright-end excess, which they associated primarily with the contribution of brightest cluster galaxies (BCGs), reinforcing the notion that the high-luminosity tail can carry an imprint of the densest environments even in nominally global measurements. \cite{2003ApJS..149..289B}, combining SDSS with 2MASS, argued that near-infrared global LFs inferred from 2MASS alone are biased against low-surface brightness galaxies and that correcting this effect yields a steeper faint-end slope and revised K-band luminosity density. Moreover, nearby-universe studies have highlighted that the global faint-end behaviour is not universal across galaxy classes: \cite{10.1046/j.1365-8711.2002.05651.x} found markedly different faint-end slopes for early-type versus late-type systems, underscoring that the aggregate global LF reflects the superposition of multiple populations whose relative abundances vary across samples and environments. Taken together with \newupdate{other} large-survey results \citep[e.g:][]{Norberg2002,  2005ApJ...631..208B, Loveday2011}, these empirical findings motivate renewed attention to how specific populations, such as BCGs at the bright end and low-surface-brightness galaxies at the faint end, can shape the global LF and related quantities such as luminosity density and stellar mass density.

Further work sought to capitalise on the large statistical advances that came with 2dFGRS and SDSS \newupdate{by revisiting the pioneering work of
\cite{1988ARA&A..26..509B} and}
 demonstrating that LF shape
and normalisation vary systematically with galactic environment, reflecting changes in the underlying galaxy population across the cosmic web. Empirically, dense environments such as groups and clusters are associated with enhanced fractions of luminous, red, early-type galaxies, while lower-density field regions host higher proportions of star-forming, late-type systems, and these differences propagate directly into environment-binned LFs. In 2dFGRS, type-dependent and environment-sensitive LF measurements \citep[e.g.][]{10.1046/j.1365-8711.1999.02721.x, Norberg2002, 2002MNRAS.333..133M} established that early-type and late-type LFs differ in both characteristic luminosity and faint-end slope, implying that shifts in the proportions of different galaxy types with environment naturally produce measurable LF variations. Subsequent SDSS analyses similarly emphasised systematic LF dependencies on galaxy colour, morphology, and star-formation activity \citep{2005ApJ...631..208B}. More targeted environmental studies have sharpened this view by explicitly comparing field, group, and cluster LFs and by connecting LF variations to physically motivated environment definitions. For example, \cite{10.1093/mnras/stw898} examined LF differences between field and denser systems (groups/clusters), while \cite{McNaught-Roberts_2014} quantified LF trends with local density, and \cite{10.1093/mnras/stx2638} explored LF variations across different cosmic-web classifications. These studies collectively support the empirical conclusion that environment affects both the abundance of luminous galaxies and the representation of faint galaxies, where trends can differ by galaxy type and may be entangled with the detectability of low-surface-brightness populations \citep[as emphasised by][]{2003ApJS..149..289B}. In parallel, other multi-band LF programs such as GAMA \citep{2011MNRAS.413..971D} have provided a consistent empirical baseline across optical filters \citep[$ugriz$; see][]{Loveday2011}, facilitating comparisons of how environment-dependent population shifts manifest with wavelength. More broadly, characterising LF variation with environment (across density, halo mass, cosmic-web morphology, etc) has become central to linking galaxy demographics to the physics of quenching, merging, and hierarchical growth. Although this paper only focuses on the global LF, the methodology in this paper will be used in a subsequent second paper where we will present density-dependent LFs.} 

\update{The global LF has also} been a useful method for constraining various galaxy formation models. 
\cite{2012MNRAS.422.2816B} discuss how in semi-analytical galaxy formation models  
\citep[e.g.][]{2012NewA...17..175B,2016MNRAS.462.3854L,2024A&A...687A..68D} the bright-end is shaped by AGN feedback while the faint end is influenced by SN feedback.
Also, \cite{10.1046/j.1365-8711.2002.05651.x} find that there is a discrepancy in the faint end slope
between simulations and empirical results, with the observed faint end of the LF being less steep than predicted. 
This result has been confirmed by several studies,
\citep{Hoyle_2005, Moretti} suggesting that there is further scope to improve galaxy formation simulations.

In addition, the bivariate and conditional LFs are another useful tool to better understand galaxy populations.
Given that it is known that the LF varies as a function of intrinsic galaxy properties, prior studies have used the bivariate LF as a means of probing this relationship by examining the luminosity conditional on other factors. For example, bivariate LFs have been constructed for the SDSS $r$-band and stellar mass, inverse concentration index, morphology, Sérsic index, and other properties \citep{2006MNRAS.373..845B}. In this paper, we use the bivariate LF to investigate the $r$-band luminosity conditional on other luminosity bands.

In recent years, the Dark Energy Spectroscopic Instrument (DESI) has greatly increased the available data for such studies \citep{DESI2016.Instr,DESI2022.KP1.Instr}.
The Bright Galaxy Survey (BGS) extends fainter than comparable studies (two magnitudes fainter than the SDSS Main Galaxy Survey) with a correspondingly higher median redshift \citep{Hahn_2023}. This provides us with an opportunity to further constrain the estimate of the LF and explore galaxy evolution. We seek to capitalise on these data to measure the LF in different bands. Additionally, we investigate LFs split by colour.

In section {\ref{Overview of DESI BGS Data}}, we describe the input catalogue including details of the redshifts and completeness weights. In section \ref{Method of Global LF estimation}, we discuss the calculation of the absolute magnitudes with the use of k-corrections and e-corrections as well as our choice of LF estimator. Section~\ref{Results of Global LF estimation} presents our LF results and we draw our conclusions in
Section~\ref{Conclusions}. Throughout this paper we adopt a standard $\rm \Lambda$CDM cosmology with $\Omega_\textrm{M} = 0.313$, $\Omega_{\rm \Lambda}=0.687$ and $H_0 = 100h\,\rm kms^{-1}Mpc^{-1}$ \citep{refId0}.\footnote{For this paper, we choose to work in units in which the exact dependence on $h$ is explicit.}

\section{Overview of DESI BGS Data}\label{Overview of DESI BGS Data}

In this section, we describe the DESI survey and the BGS data. We provide details of the galaxy catalogue in Section~\ref{sec:input_cat}. In particular, we discuss the LSS  catalogues that are described in \cite{2025JCAP...01..125R}. We further describe in Section~\ref{sec:redshifts} how DESI determines the spectroscopic redshift for each object. In Section~\ref{weights}, we describe how we correct for target and redshift incompleteness in the survey. In section~\ref{sec:randoms} we describe the random catalogues that match the \cite{2025JCAP...01..125R} LSS catalogues.

\subsection{Input Catalogue} \label{sec:input_cat}

The Dark Energy Spectroscopic Instrument (DESI) 
is a next-generation Stage IV survey based at Kitt Peak, Arizona as part of the 4-m Mayall Telescope. 
The primary goal of DESI is to map the large scale galaxy and quasar distribution in order 
to provide sub-percent precision measurements of a range of cosmological parameters. 
This is achieved by observing targets using spectrographs fed by almost 5000 robotically positioned optical fibres 
across a $3^\circ$ field of view
\citep{Spectro.Pipeline.Guy.2023,SurveyOps.Schlafly.2023,FocalPlane.Silber.2023,Corrector.Miller.2023,FiberSystem.Poppett.2024}.
As part of this project, there are five main samples of data: Emission Line Galaxies (ELGs), Luminous Red Galaxies (LRGs), Quasars (QSOs), the Milky Way Survey (MWS) and the Bright Galaxy Survey (BGS) \citep{desicollaboration2016desi}.

BGS is the component of DESI that focuses on the mapping of more than 10 million galaxies from $0 < z < 0.6$. BGS Bright is a $r \lesssim 19.5$ magnitude limited sample, while BGS Faint covers the fainter range $19.5 < r < 20.175$ but additionally has a colour-dependent fibre-magnitude limit  \citep{Hahn_2023} to ensure the redshift measurements have a high success rate. 
Here we limit ourselves to the BGS Bright sample. DESI targets, including those in BGS, are selected based on applying photometric criteria to the photometric data taken with the DESI Legacy Surveys, BASS/MzLS in the North and DECaLS in the South \citep{Dey_2019}. The target selection procedure for BGS is described in detail in \cite{Omar2020}, while the final target selection choices are summarised  in \cite{Myers_2023}. As a result of the differing photometry in North and South, the DESI BGS Bright survey is magnitude-limited to $r< 19.54$ in the North (dec > 32.375 deg) and $r < 19.5$ in the South to achieve the same surface density of targets in both hemispheres. These are extinction-corrected apparent magnitudes which are based on the SFD dust map \citep{Schlegel_1998}. BGS Bright - which will be the focus of this paper - is similar to the depth of the Galaxy And Mass Assembly (GAMA) survey \citep{2011MNRAS.413..971D}, 
%and two magnitudes fainter than the SDSS Main Galaxy Survey \citep{2002AJ....124.1810S}, % we already made this point earlier and the following much larger area does apply compared to SDSS
but covers a much larger area of the sky\footnote{This excludes low completeness single pass regions around the edge of the survey and a recently added small extension increasing the overlap with the
DES survey. If both were included the final area would be closer to 15,000~deg$^2$.}
at 14,000~deg$^2$.

\update{
The DESI magnitudes are total model magnitudes computed by the Legacy Survey pipeline 
\citep{Dey_2019}. Each detected source is fitted by each of the following seeing convolved model profiles: a point source (PSF),  a round exponential (REX), elliptical exponential (EXP), elliptical de Vaucoleurs (DEV) and elliptical Sérsic (SER). The $\chi^2$ of the fit, taking into account a penalty for the number of parameters in the model, is used to determine the best fit model. The total magnitude is determined by the integral of the total flux of this best-fit model.
 DESI fibre magnitudes are defined as the prediction of the best fit model for a 1.5 arcsec diameter fibre aperture and assumed fiducial seeing of 1 arcsec FWHM.
We later apply distance modulus, passband and evolutionary corrections to these magnitudes to define rest-frame absolute magnitudes, as discussed in Section~\ref{k-corrections and e-corrections} and \ref{e-corrections}.}

In this paper, we make use of DESI DR2 data 
(otherwise known as Year Three (Y3) data). This is the data of the main DESI survey collected in the first three years of the survey, with key results released in 2025 \citep{desicollaboration2025datarelease1dark, Anonymous_2025, tr6y-kpc6}.

%after approximately three years of operations, with key DESI results released in 2024 \citep{2024arXiv240403000D,2024arXiv240403002D}. 

Our catalogue contains a total of 7,874,903 galaxies spanning an area of $12,355.08~\rm{deg}^2$, shown in Fig. \ref{fig:DESI_map}. In addition, the DESI Y3 catalogue also makes use of data from the Wide-Field Infrared Survey Explorer (WISE). WISE is a space-based infrared survey which mapped the whole sky with multiple passes \citep{2010AJ....140.1868W}.  In particular, we use magnitudes from the $w1$ band centred at 3.4$\mu$m with an angular resolution of 6.1", in order to present a $w1$-band LF. 

\subsection{Redshifts} \label{sec:redshifts}
In order to determine the redshifts of galaxies from the measured spectra, DESI makes use of Redrock \citep{2024AJ....168..124A}, a template-based classifier which classifies spectra as GALAXY, STAR or QSO and assigns a redshift to each spectrum based on $\chi^2$ minimisation of a linear combination of Principal Component Analysis (PCA) basis spectral templates.  Although the PCA-based redshift estimator can occasionally yield unphysical models, such as negative emission lines, Redrock mitigates this issue reasonably well by applying penalties \cite[see][for more details]{2024AJ....168..124A}. Redrock templates were constructed using eBOSS spectra. Redrock additionally generates a $\Delta \chi^2$ value which is the difference in $\chi^2$ between the best and second-highest likelihood peaks with different redshifts, which acts as a metric of the confidence of the redshift determination. 

A large number of spectra were evaluated by visual inspectors to assess the validity of the Redrock algorithm, as detailed in \cite{Lan_2023}. For the Survey Validation (SV), BGS Bright had an average assessed Visual Inspection Quality $> 2.5$\footnote{A visual inspection quality of 3 corresponds to a  probable classification with at least one secure spectral feature and continuum or many weak spectral features, e.g., spectra with a strong emission line feature and weak Balmer series absorption lines while quality 2 corresponds to a classification with one strong but unidentified spectral feature \citep{Lan_2023}.} 
for 99.6\% of galaxies based on 1037 sources, indicating that Redrock returns robust redshifts for the vast majority of objects. Moreover, \cite{Lan_2023} investigates the use of $\Delta \chi^2$ as a metric for assessing the reliability of the Redrock redshifts. For BGS Bright, the authors find 100\% purity for $\Delta \chi^2 > 40$ in the VI sample of 2718 BGS target galaxies, indicating that this is a useful threshold for ensuring that sources have the correct redshifts.

\update{
The redshifts returned by Redrock are relative to the solar system barycentre. For the purposes of our analysis we transform these redshifts into the Cosmic Microwave Background (CMB) rest frame and correct for local peculiar velocities using the CosmicFlows-4 grouped flow field \citep{2023A&A...670L..15C}.
The effect of these transformations on our luminosity function estimates is small but moderately significant given the very small statistical errors. The bright end of our luminosity functions are dominated by galaxies at $z>0.1$  where the
change in distance modulus is insignificant. At lower redshift, the change in distance modulus for individual galaxies is more significant 
but the wide sky coverage of DESI combined with variation of the flow field across the sky results in little net change, even for the faint end of the luminosity function.
}

\subsection{Weights} \label{weights}

To construct luminosity functions, it is important to correct for incompleteness within the DR2 dataset, including systematic effects in the input catalogue, target incompleteness and redshift success rates.

Typically, a systematic weight (WEIGHT\_SYS) is used to account for target density fluctuations due to imaging conditions and foregrounds. Specifically, this weight corrects for unphysical correlations of the target density with dust extinction, stellar density, and HI maps (DESI Collaboration, private correspondence). We observe that these weights are close to unity and make a negligible difference to our results. In particular, they have no impact at all on the faint-end of the $r$-band LF. Their value is in correcting for large scale spatial variations in the completeness, not in determining global quantities such as the LF. As such, we have chosen to ignore this systematic weight in this paper.
\update{However in addition to these systematics incompleteness at low surface brightness could 
be important at the faint end of the galaxy luminosity function. We therefore explicitly investigate this in Section~\ref{sec:LFs}.}

An additional source of incompleteness is targeting incompleteness, which in this case is a correction factor to account for targets that were not observed. In tiles for which there has only been a single pass, there will be targeting incompleteness as only one object can be targeted per fibre - specifically for each fibre, this target will be within the unique patrol region of each fibre. This means that in regions of high target density, not all targets will be assigned to fibres. Furthermore, fibres cannot be placed arbitrarily close to each other due to the physical and mechanical constraints of the fibre positioners. This limit can also lead to observable targets not being assigned a fibre. The footprint of the survey is determined by the area of sky that is the union of sky reachable by good fibres on observed fields. DESI conducts multiple passes to reduce targeting incompleteness. However, in regions covered by multiple tiles, the targeting incompleteness is reduced but is rarely completely removed. To correct for this incompleteness, a weight,
 $w_\textrm{comp}$ (WEIGHT\_COMP), is defined for DR2 in \cite{2025JCAP...01..125R}. In particular, $w_\textrm{comp}$ accounts for the observable targets that were not assigned a fibre by assigning their weight to a neighbouring object that was assigned a (working) fibre.

In addition, not all targeted galaxies may receive redshifts (e.g: due to a failure of Redrock to fit a model spectra with confidence). Whilst the DESI LSS catalogues do calculate a redshift completeness weight (WEIGHT\_ZFAIL), this weight is designed to correct to a uniform sample (with uniform incompleteness) rather than to a complete sample. As such, this weight is adequate for clustering but less useful for LFs. We choose instead to make our own direct estimate of the redshift completeness weights ($w_\textrm{z}$).

To do this, we take the full LSS catalogue, that is, the LSS catalogue that contains all potentially observed targets within the footprint (see Section 2.3 of the KP3 paper and Section 4 of Ross et al. 2025). From this, we then define a `zgood' subset (i.e. the standard LSS catalogue) by applying following selection cuts:
\begin{enumerate}
\item DELTACHI2 ($\Delta \chi^2$) > 40.
\item ZWARN = 0
\item 0.002 < z < 0.6
\end{enumerate}

Here ZWARN is a bitmask that incorporates various information to indicate if the redshift has problems, such as fibre problems, low wavelength coverage, spectrum fitting problems, etc. ZWARN=0 indicates that there are no apparent problems with the redshift. The criteria above ensure that the redshift is reliable and excludes the redshift range contaminated by stars. In addition, the upper redshift cut ensures that we avoid objects with potentially spuriously fitted spectra. We note that $99.55\%$ of the data is $z < 0.6$, so this excludes relatively few galaxies. 

The overall redshift completeness of the survey is simply the number of objects in `zgood' over the number in the full observed catalogue. We calculate the incompleteness in bins of $r$-fibre magnitude and the template signal to noise squared for BGS \cite[TSNR2$\_$BGS, defined in][]{Spectro.Pipeline.Guy.2023}. 
The TSNR2$\_$BGS is the square of the expected spectral signal-to-noise ratio for a fiducial BGS source for the specific fibre and observation, as a result it is proportional to the effective exposure time for the individual target.\footnote{Specifically, the effective exposure time ($T_{\rm spec}$) is related to TSNR2$\_$BGS as per the following relation: $T_{\rm spec} = (0.135 \rm{sec}) \times \rm{TSNR}^2_{\rm BGS}$
\citep{2025JCAP...01..147K}.
}
Hence
%[TSNR2BGS is proportional to the mean squared difference between a BGS template spectrum and the median filtered version of that same spectrum. It is also inversely proportional to the flux uncertainty. As a result, TSNR2BGS is proportional to the effective exposure time.] 
we select $r_\textrm{fibre}$ and TSNR2$\_$BGS as variables for this calculation as we expect redshift completeness to decrease for fainter $r_\textrm{fibre}$ and smaller TSNR2$\_$BGS.

\begin{figure}
    \centering
    \includegraphics[width=\columnwidth]{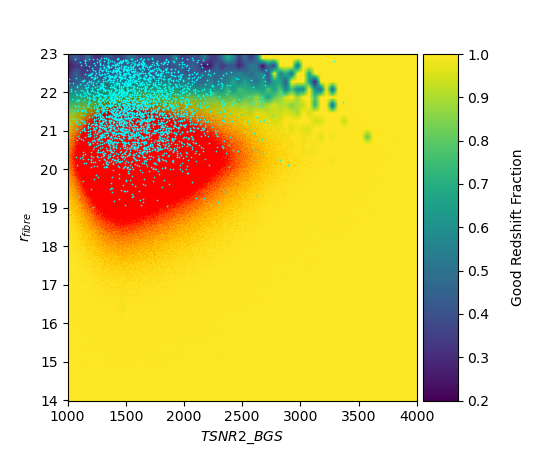}
    \caption{\newupdate{The empirical redshift completeness as a function of the fibre magnitude, $r_\textrm{fibre}$, and the TSNR2$\_$BGS with different galaxy samples overplotted. TSNR2$\_$BGS is the expected spectral target signal-to-noise ratio for a fiducial BGS source. This plot is for the South region.
    All the galaxies in this region are plotted as red points with low opacity. The majority fall where the redshift completeness is very high. The cyan points are a subset of faint galaxies with absolute $r$-band magnitude fainter than $-13$ (see the discussion of modelling in the incompleteness in this regime in Section~\ref{sec:LFs}).}
    To define a weight, $w_{\rm z}$, to correct for this incompleteness we interpolate the binned incompleteness using the Cloud-In-Cell technique and
    take its inverse. Note that for empty pixels, we default to a value of 1.}
    \label{fig:weight_lookup}
\end{figure}

For the North region of the survey the redshift incompleteness is shown in Fig.~\ref{fig:weight_lookup}.
We use the Cloud-In-Cell interpolation technique \citep{1988csup.book.....H} to smooth over the bins and
define the redshift incompleteness at any arbitary 
 $r_\textrm{fibre}$ and  TSNR2$\_$BGS and
 define its inverse as the redshift completeness weight, $w_{\rm z}$. 

The total weight applied to the $i^\textrm{th}$ galaxy to correct for both targeting and redshift incompleteness is the product of these two weights
%We confirm that our total weights are higher on average than the total weight (`WEIGHT') provided in the DESI catalogue.
\begin{equation} \label{eq:weights}
    w_i = w_{\textrm{comp},i} \, w_{\textrm{z},i}.
\end{equation}

For the target incompleteness weight, we emphasise that the values are all integers, with a minimum value of 1.0. As such, the median target incompleteness weight is 1.0, with 97.6\% of weights $\le 3.0$. 

For the redshift completeness weight, the overall median is 1.0015 with the 10th percentile 1.0000(2) and the 90th percentile 1.0173. If one looks at the
faintest objects the weights are more significant. 
For the 2.33\% of objects with $r_{\rm fibre} > 21.5$, the redshift incompleteness weight median is 1.1626 with the 10th percentile 1.0821 and the 90th percentile 1.6299. 

For all objects, the median total weight is 1.0026 with the 10th percentile 1.0000(3) and 90th percentile 2.0074. For the objects that have $r_{\rm fibre} > 21.5$, the median total weight is 1.2022 with the 10th percentile 1.0852 and 90th percentile 2.5776.

\subsection{Random Catalogue} \label{sec:randoms}

To define the selection function of the survey a catalogue of randomly positioned points is generated over the DESI sky footprint with a number density of 2500 objects per square degree, as described in \cite{2025JCAP...01..125R}. Then, only the randoms that are reachable by a good fibre of an observed tile are kept in the catalogue. These randoms are additionally assigned redshifts and other galaxy properties, with the assignment done randomly from the selected galaxy sample. It should be noted that North and South property assignment is done separately from each other to allow for the difference in selection functions. As described in Section~\ref{weights}, DESI assigns fibres to obtain the spectra of galaxies. This means that a single pass will suffer from target incompleteness as there will be `holes' in the survey, due to the gaps between petals, holes from where the Guiding, Focusing and Alignment (GFA) cameras are, and malfunctioning fibres. The distribution of the randoms fully takes this into account. We use the total number of randoms to quantify the total area of the sky observed, but also their distribution to map this incompleteness \citep[see figures~2 and~3 in][]{2025JCAP...01..125R}.

\begin{figure}
    \centering
    \includegraphics[width=\columnwidth]{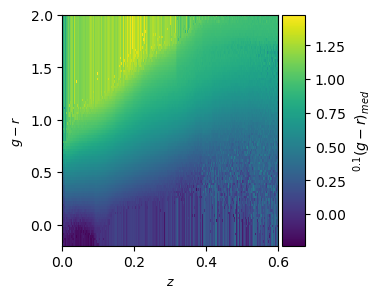}
    \caption{The median rest-frame $^{0.1}(g-r)$ colour of galaxies from the FSF South catalogue in bins of redshift and observer-frame colour. This provides a look-up table to infer the rest-frame colour from the observed properties. A separate lookup table is used for the North.}
    \label{fig:col_lookup}
\end{figure}

\section{Method of Global LF estimation} \label{Method of Global LF estimation}

\subsection{k-corrections}
\label{k-corrections and e-corrections}

Typically, absolute magnitudes are corrected by k-corrections to account for band-shifting effects, specifically that the observed passband will map to different rest-frame passbands for galaxies at different redshifts \cite[see][for a comprehensive overview of k-corrections]{hogg2002k}. In order to compare the photometric properties of galaxies at different redshifts, we need to transform their photometry to that of a fixed combination of reference frame and filter curve. For ease of comparison, we have chosen this reference frame to be the SDSS $r$-band (and $g$-band) filter curves with a reference redshift of $z_\textrm{ref}=0.1$ as adopted by \cite{Zehavi_2005} and \cite{Loveday2011}. Our goal is to model the mean redshift dependent k-correction in each band and its dependence on a single rest-frame colour.

%In order to take into account the differing photometry of DESI, we require k-corrections for each individual galaxy. 
To compute absolute magnitudes and k-corrections FastSpecFit\footnote{The documentation for FastSpecFit can be found at https://fastspecfit.readthedocs.io/en/latest/} (FSF) was developed to perform fast spectral synthesis and emission-line fitting of DESI spectra and broadband photometry \citep{2023ascl.soft08005M}. In particular, FSF works by simultaneously fitting model SEDs to a combination of the broadband photometry and the aperture-corrected DESI spectral photometry. From these model SEDs, FSF absolute magnitudes and k-corrections have been calculated. We require full redshift-dependent k-correction functions so that we can calculate $V_\textrm{max}$ \footnote{$V_\textrm{max}$ is the volume within which the galaxy can be re-positioned and still satisfy all the selection criteria to be included in the sample that is being analysed. For instance, there is a maximum redshift to which the galaxy could be relocated before its apparent magnitude is too faint for it to be included in the sample and this depends on the k-correction at this redshift.} as defined in Eqn. ~\ref{eq:vmax}. Whilst FSF is able to do this for each individual galaxy, for convenience and computational speed we instead construct  polynomial k-corrections fitted to 
the FSF catalogue of k-corrected magnitudes. Our method is described below.

First, we create a rest-frame colour lookup-table, where we generate a 2D histogram of observer-frame $g-r$ colour against redshift and compute the median rest-frame FSF colour in each pixel. (Fig.~\ref{fig:col_lookup}). Using this table, and cloud-in-cell interpolation, each galaxy in the Y3 DESI catalogue is assigned the median rest-frame colour corresponding with its observed colour and redshift.
This assigned colour is then used to bin the galaxies into 7 rest-frame colour-bins each containing an equal number of objects.

\begin{figure}
    \centering
    \includegraphics[width=0.95\columnwidth]{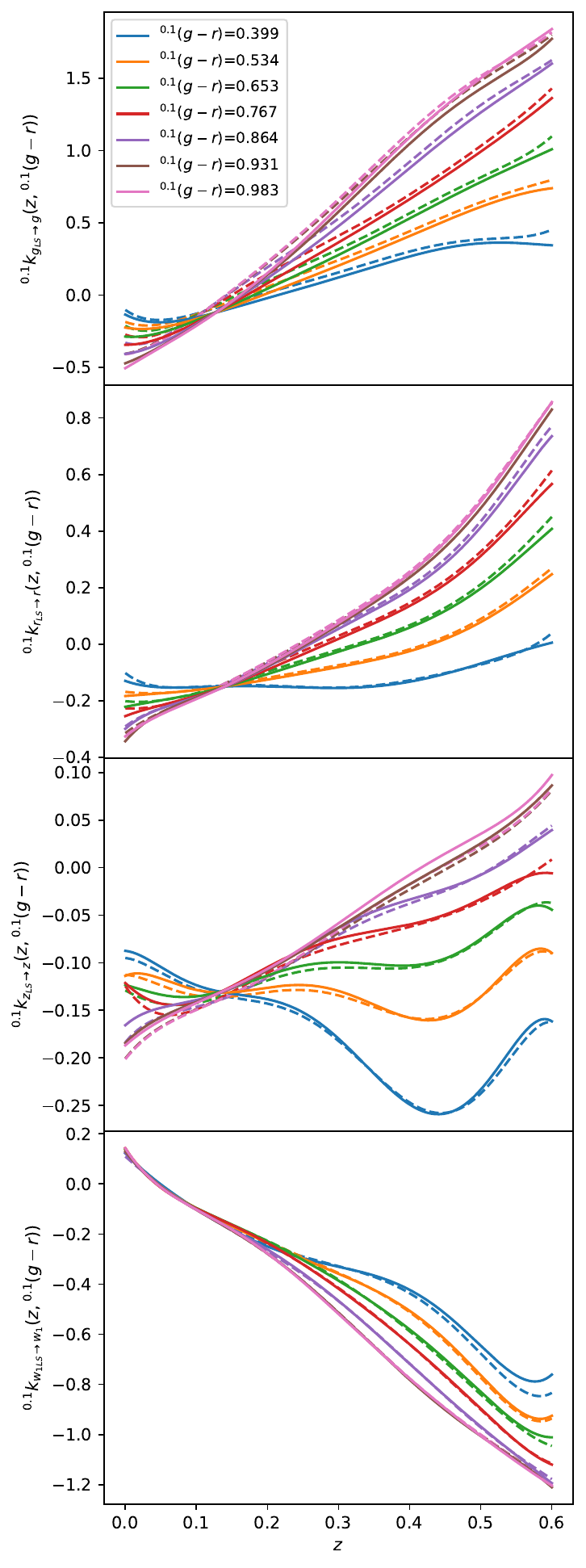}
    \caption{The k-correction polynomials to the SDSS $g$, $r$, $z$ and WISE $w1$ bands with $z_{\rm ref} = 0.1$ from the respective observer frame DECaLS band (South, solid line), BASS/MzLS band (North, dashed line) and WISE bands. The $r$, $z$ and $w1$-band k-corrections are direct fits to the FSF data. For the $g$ band, the $r$-band polynomials are transformed to the $g$ band using Eqn. \ref{eq:k-corr_eq} and the figure shows polynomial fits to the resulting $g$-band k-correction.}
    \label{fig:kcorr}
\end{figure}

Each colour-bin is split into a range of redshift-bins and the median FSF k-correction is calculated in each bin. \footnote{The FSF $r$-band k-corrections as provided have an intentional discontinuity at $z \sim 0.3$ due to switching the reference band. In order to standardise the FSF k-corrections to be from the BASS/DeCALS $r$-band to the SDSS $r$-band across all redshifts, we formally use a `derived' FSF k-correction, calculated from the FSF provided absolute magnitude, the apparent magnitude and the redshift (using Eqn.~\ref{eq:mags}). We also use this methodology for the $g$, $z$ and $w1$ bands.} A least-squares polynomial fit is then performed to these median $r$-band k-corrections to find a 7th-order polynomial for each colour bin. The North and South are modelled separately (second panel of Fig.~\ref{fig:kcorr}). From this plot, we see that the North and South curves are slightly different reflecting the differences in the North/South photometry. Furthermore, we note that there is a pinch point at about $z=0.14$ which corresponds to the redshift at which the central wavelength of the DESI $r$-band filter best matches with that of the SDSS $r$-band filter at a reference redshift $z_{\textrm{ref}}=0.1$. This corresponds to the 3.74\% shift in the effective wavelength between the SDSS $r$-band filter (6205.83Å) and the BASS $r$-band filter (6437.79Å), as  $[(1+0.141) / (1+0.1)]-1$ is approximately 3.73\%.

We note that there is a slight offset in the pinch points for North and South in the $g$-band in a way that is not apparent in the $r$, $z$ bands. This is explained by the BASS and DECaLS $g$-band filters being significantly different, with a greater difference than in any other band. Moreover, the $w1$-band k-correction is simply the native $w1$ filter shifted to the reference redshift. As such, it has a pinch point consistent with $k_{w1}(z=0.1) = -2.5 \log(1+0.1)$, where we note that the very slight deviation from this can be attributed to the polynomial fits not being constrained to go through this point.

We conducted additional tests to ensure that these k-correction polynomials remain largely invariant to choices in colour and redshift bin size - for example, choosing a larger number of colour-bins for k-correction polynomial fitting has a small but ultimately negligible impact on the LF compared to the jackknife error. .

When assigning a k-correction based on redshift and observer-frame colour, we use a cubic interpolation scheme between the 7 polynomials. From this, with the BGS catalogues we provide %$\zref{M_r} - 5\log h$, the 
absolute magnitude in the SDSS $r$-band with reference redshift $z_{\textrm{ref}}=0.1$, defined by

\begin{multline} \label{eq:mags}
    \zref{M}_{r} - 5 \log_{10}h = m_{r_{\rm LS}} - 5\log_{10}\left(\frac{d_L(z)}{h^{-1} \rm Mpc}\right) - 25 \\ - \zref{k}_{r_{\rm LS} \rightarrow r}(z,\zref{(g-r)}) - \zref{E}(z).
\end{multline}

Here, the subscript $r$ represents the SDSS $r$ band,
while subscript $r_{\rm LS}$ represents the Legacy Survey $r$-band either from DeCALS in the South or BASS in the North.
$\zref{k_{r_{\rm LS} \rightarrow r}}(z, \zref{(g-r)})$ represents our fitted polynomial 
k-correction that takes us from Legacy Survey observer frame $r$-band to the rest-frame SDSS $r$-band with a reference redshift of $z_\textrm{ref}=0.1$, where $^{0.1}(g-r)$ is the rest-frame colour. 
$d_L(z)$ is the luminosity distance to the redshift $z$, determined using our assumed cosmology. Optionally and in addition, an e-correction may be applied in order to account for the intrinsic luminosity evolution of a galaxy over time, which is represented by the $\zref{E(z)}$ term; this is discussed in Section \ref{e-corrections}. 

For the $w1$ and $z$ bands, we construct polynomial fits to the median FSF k-corrections in bins of $^{0.1}(g-r)$
in just the same was as we do in the $r$ band.
To set the $g$-band absolute magnitude
we use 
$$
\zref{M_g} = \zref{M_r} + ^{0.1}(g-r)_{\rm med}
$$
where $^{0.1}(g-r)_{\rm med}$ comes from the look-up table.
This implies the $g$-band k-correction we use is given by
\begin{multline} \label{eq:k-corr_eq}
\zref{k_{g_{\rm LS}\rightarrow g}(z,\zref{(g-r)})}
=  
\zref{k_{r_{\rm LS}\rightarrow r}(z,\zref{(g-r)})}\\
+(g-r) - \left(\zref{M_g} - \zref{M_r}\right)
\end{multline}
and its median in colour bins is shown in the top panel of Fig.~\ref{fig:kcorr}
\footnote{Along with our luminosity function estimates the
polynomial coefficients of our k-correction fits are available in electronic form at
\url{https://icc.dur.ac.uk/data/}}.

We emphasise that whilst this methodology of polynomial fitting is similar to that described in \cite{Adame2024}, those k-correction polynomials are based on GAMA DR4 data, while in this paper we have updated the method to make use of a colour lookup table and to directly use DESI BGS galaxies in order to account for the DESI photometry and its differences in the North and South regions.

\begin{figure}
    \centering
    \includegraphics[width=\columnwidth]{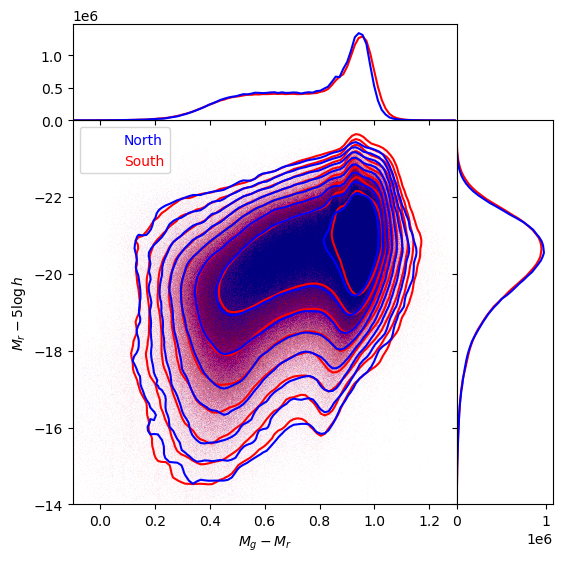}
    \caption{Rest-frame colour vs. absolute magnitude distributions in North and South.
    Contours are plotted representing number density, with each successive contour a factor of 2 larger in number density (starting at 160 objects per bin). The top histogram shows  the $^{0.1}(g-r)$ colour distribution for North and South. The right histogram shows the $r$-band absolute magnitude distribution. Both histograms have been normalised by sky-area to adjust for the fact that South is approximately double the size of North. }
    \label{fig:col_mag}
\end{figure}

Fig. \ref{fig:col_mag} presents the $\zref{(g-r)}$ rest-frame colour distribution and the $r$-band absolute magnitude distribution. We observe that there exists a slight offset in the North and South colour distributions. We attribute this to a possible small error in the red end of one of the filter curves. We note that if we make an empirical correction for this by shifting the $r$-band magnitudes of the galaxies in the North such that we preserve their ranking in rest-frame colour and then exactly match the cumulative rest-frame colour distribution of the South, then this makes a negligible difference to our subsequent results (see Section~\ref{Results of Global LF estimation}).

\subsection{e-corrections} \label{e-corrections}

We follow other authors and implement an additional e-correction in the calculation of absolute magnitude, and hence model the evolution of the galaxy LF as being caused by the evolution of the individual galaxy luminosities.
We follow the convention of \cite{McNaught-Roberts_2014} 
and model the redshift dependence to first order as 
\begin{equation} \label{eq:e-corr}
    ^{z_{\rm ref}}E(z) = -Q (z-z_\textrm{ref})
\end{equation}
with one free parameter $Q$.
 Other papers \cite[e.g.][]{Loveday2011} additionally allow the galaxy population to evolve in number density using a second parameter $P$ as 
 
\begin{equation}
P(z) = P(z_0) 10^{0.4P(z-z_0)}    
\end{equation}
 
 \cite{Loveday2011} find that for all bands and colour samples, the estimates of $P$ and $Q$ are strongly anti-correlated. 
 To avoid this degeneracy we simply assume $P=0$ (no density evolution). One advantage of doing this is that it allows for a more direct comparison to the results in \cite{McNaught-Roberts_2014} who also fixed $P=0$. This was useful in our initial stages when validating our methodology on GAMA data.

\update{In the $r$-band, to constrain $Q_r$} we consider the distribution
of the ratio of $V/V_\textrm{max}$ -- where $V$ is the survey volume up to the observed redshift of the galaxy
and $V_\textrm{max}$ is the maximum volume over which the galaxy could be seen, given the observed properties of the galaxy and the apparent magnitude limits of the survey. $V_\textrm{max}$ is defined 
by
\update{
\begin{equation} \label{eq:vmax}
    V_{\textrm{max} }(L^r_i,k_i) = \frac{4}{3} \pi f_{\rm sky}[d(z_{\textrm{max,}i})^3 - d(z_{\textrm{min, } i})^3] ,
\end{equation}
where $z_{\textrm{min}, i}$ and $z_{\textrm{max,} i}$ are the minimum and maximum redshifts respectively at which a galaxy could be observed given its 
$r$-band luminosity, $L^r$, its colour dependent k-correection, $k_i$, and the redshift limits of the BGS sample.} $f_{\rm sky}$ is the fraction of the sky area over which the North or South region of the survey is defined. For the North, $f_{\rm sky} = 0.09278$ and for the South, $f_{\rm sky} = 0.20671$.

\begin{figure}
    \centering
    \includegraphics[width=\linewidth]{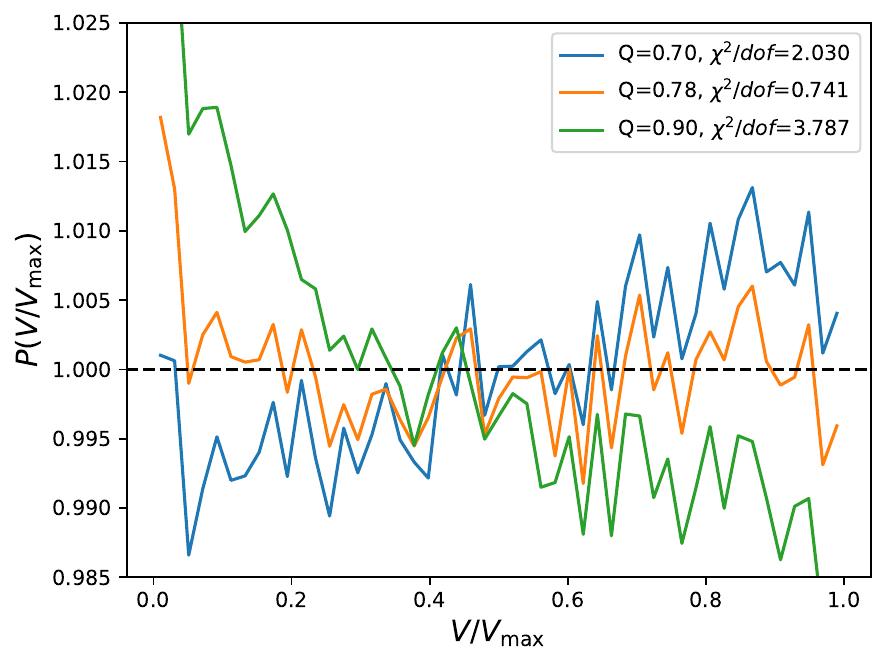}
    \caption{The \update{North} $V / V_{\rm max}$ distribution for $Q_r=0.78$ (the optimal value we find) and a number of nearby values of $Q_r$. The reduced $\chi^2$ value is presented for each $Q_r$-value. %\todo{Can the $Q$s in the legend be changed to $Q_r$?}
    }
    \label{fig:e_corr_vmax}
\end{figure}

\update{If the chosen value of $Q_r$ correctly models the luminosity evolution, then the $V/V_{\rm max}$ distribution should be uniform, aside from small fluctuations caused by large‑scale structure.
We therefore vary $Q_r$ to minimise the $\chi^2$ difference between the measured and ideally uniform $V/V_{\rm max}$ distributions, using jackknife errors (see Section~\ref{Luminosity Functions}).
The resulting $V/V_{\rm max}$ distributions for several values of $Q_r$ in the North are shown in Fig.~\ref{fig:e_corr_vmax}; we obtain very similar results for the South. From this procedure, we find a global value of $Q_r=0.78\pm 0.20$, compared with $Q_r=0.97\pm 0.15$  reported by \cite{McNaught-Roberts_2014}. Their differing statistical method and redshift range mean that exact agreement is not expected.

Because DESI BGS is selected in the $r$-band, only $Q_r$  is required to compute $V/V_{\rm max}$ for each galaxy. However, $Q_r$ may not fully describe the luminosity evolution in the other bands. We therefore allow the value of $Q$ used when defining absolute magnitudes in the $g$, $z$, and $w1$ bands via Eqn.~\ref{eq:e-corr} to vary independently. To constrain these parameters, we divide the catalogue into two samples of roughly equal size above and below $z=0.2$. In each band, we vary $Q$, compute the luminosity function in each redshift slice, and determine the value of $Q$ that minimises the $\chi^2$ difference between the two LFs over a chosen magnitude interval. This interval spans two magnitudes and begins just brightward of the faintest magnitude at which the high‑redshift sample is complete: $[-19.75, -21.75]$ in the $g$-band and $[-20.75,-22.75]$ in the $z$ and $w1$-bands. The resulting best‑fit evolution parameters are $[Q_g,Q_z,Q_{w1}]=[0.91\pm 0.20,0.45\pm 0.15,0.7\pm 0.15]$.

The evolution of the galaxy luminosity function is undoubtedly more complex than can be captured by a single-parameter model. In reality, the shape of the LF is expected to evolve with luminosity due to changing galaxy populations and the effects of mergers. Nevertheless, this simple parametrisation is sufficient for our purposes: it enables a robust measurement of the luminosity functions at our fiducial redshift of $z=0.1$.}

We additionally consider the idea of splitting the galaxies into two colour classes  (red and blue) and finding corresponding $Q$-values for each population. To define the appropriate colour-cut, we make use of the bimodal nature of the $\zref{(g-r)}$ rest-frame colour histogram, which has two distinct populations that can each be characterised by Gaussian distributions (see topmost panel of Fig.~\ref{fig:col_mag}). By fitting Gaussians to each distribution with an Expectation-Maximisation (EM) algorithm, we can split the populations based on the intersection of those two fitted Gaussians. This yields a colour threshold of $\zref{(g-r)} = 0.75$. Alternatively, we can visually identify a difference in the two populations by splitting the $\zref{(g-r)}$ histogram into different $^{0.1}M_r - 5\log h$ absolute magnitude bins as there is a region where the trough of the combined histogram tends to be very similar. This method is consistent with the result found with the EM method. We use the $V/V_\textrm{max}$ method described above and find $Q_{\rm red} = 0.23 \pm 0.3$ and $Q_{\rm blue} = 1.59 \pm 0.2$. Our value of $Q_{\rm blue}$ is larger than that of $Q_{\rm red}$, meaning that the blue population galaxies exhibit more substantial luminosity evolution than the red population. This trend is seen in other papers, notably \cite{McNaught-Roberts_2014} and \cite{Loveday2011}.

\subsection{LF Estimators} \label{Luminosity Functions}

With the absolute magnitudes and weights found for each BGS galaxy in our sample, it is now possible to measure the LF. There are numerous different methods to measure the LF \citep{1988MNRAS.232..431E, 2011MNRAS.416..739C}. In this paper, we focus on the $V_{\rm max}$ estimator as outlined in \cite{1968ApJ...151..393S}. We compare this result to other methods in Appendix \ref{appendixLFmethods}, and find that different methods yield extremely similar results. The $V_{\rm max}$ LF estimator is

\begin{equation}
    \phi(L)dL = \sum_{i=1}^{N} \frac{w_i W(L-L_i)}{V_{\rm max}(L_i,k_i)},
\end{equation}
where $V_{\rm max}(L_i,k_i)$ is defined in Eqn. \ref{eq:vmax},
 $w_i$ represents the combined weight defined in Eqn. \ref{eq:weights}, and $W(L-L_i)$ represents a binning function
\begin{equation}
    W(L-L_i) = \Theta(L_i - L + dL/2) - \Theta(L+dL/2-L_i)
\end{equation}
where $\Theta$ is the Heaviside step function.

\update{For our bivariate LF, we adapt Eqn.~\ref{eq:vmax} into the following,
\begin{equation}
    \Psi(L^r, L^g) dL^r dL^g = \sum_{i=1}^{N} \frac{w_i W(L^r-L^r_i) W(L^g-L^g_i)}{V_{\rm max}(L^r_i, k_i)},
\end{equation}
where $L^r$ and $L^g$ are the luminosities in the $r$ and $g$ bands respectively.}

By weighting each object by the inverse of its maximum detection volume, this method corrects for the issue that intrinsically faint objects are detected only in a small volume, leading to the preferential detection of intrinsically bright objects. Additionally, the $1/V_{\rm max}$ estimator has the advantages of not assuming a functional form and automatically having the correct normalisation. The disadvantage of this estimator is that it assumes all sources follow a uniform spatial distribution. This can result in distortion in the case of overdense or underdense regions \citep{1988MNRAS.232..431E}. We find that this only affects the $r$-band LF fainter than $^{0.1}M_r - 5 \log h> - 14$ (see Section~\ref{Results of Global LF estimation}); this is because of the large volume probed by BGS at these magnitudes.

We calculate both Poisson errors and jackknife errors for our LFs. The Poisson error estimate is given by 
\begin{equation} \label{eq:poisson}
    \frac{\Delta \phi(L)}{\phi(L)}  = \sqrt{\frac{\sum_i(w_i^2) W(L-L_i)}{(\sum_i w_i W(L-L_i))^2}}.
\end{equation}
We note that if all weights are unity, then this expression reduces to the standard ${1}/{\sqrt{N(L)}}$ term, where $N(L)$ is the number of objects in the luminosity bin.

The jackknife errors are calculated by splitting the footprint of the survey into 3 by 3 ($N=9$) equal area regions in the North, and 4 by 5 ($N=20$) equal area regions in the South. We have explicitly checked that using more jackknife regions does not significantly alter our error estimates.
We note that the number of jackknife regions used is different for North and South so that the regions are approximately equal in area. The formula for the jackknife error is 
\begin{equation} \label{eq:jackknife}
    \textrm{Var}(x) = \frac{\Delta \phi(L)}{\phi(L)} = \frac{N-1}{N} \sum_{i=1}^{N} (x_i - \bar{x})^2 ,
\end{equation}
where $x_i$ is the value of $\phi(M)$ calculated for the area excluding the $i^{\textrm{th}}$ region \citep{10.1111/j.1365-2966.2009.14389.x}.

\section{Results of Global LF estimation} \label{Results of Global LF estimation}

\begin{figure*}%
{%
  \label{fig:Total_scatter}%
  \includegraphics[width=0.46\textwidth]{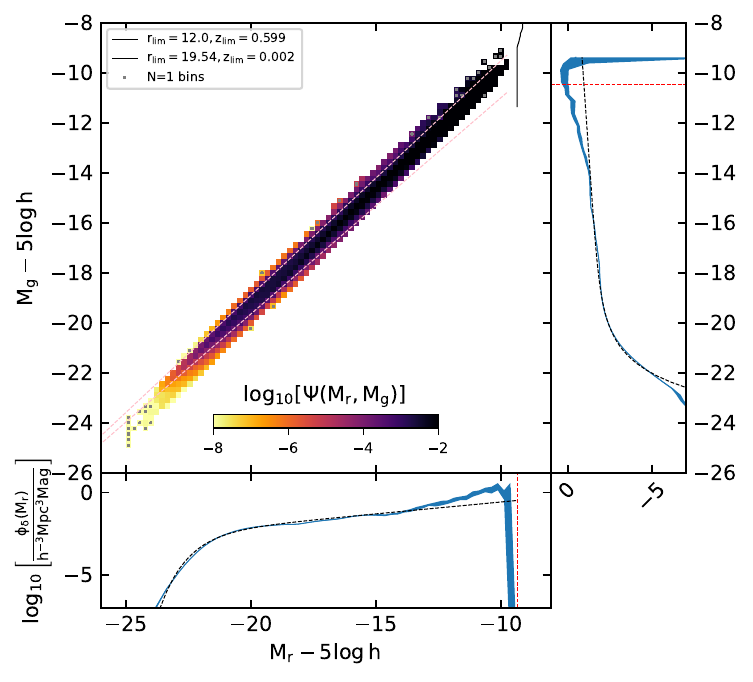}%
}%
{
  \label{fig:Num_scatter}%
  \includegraphics[width=0.46\textwidth]{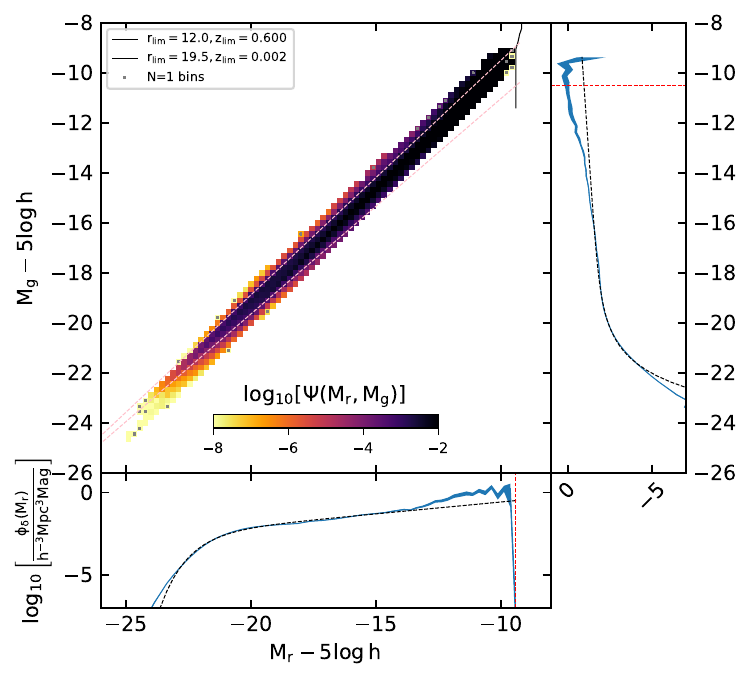}%
}%
{
  \label{fig:Raw_scatter}%
  \includegraphics[width=0.46\textwidth]{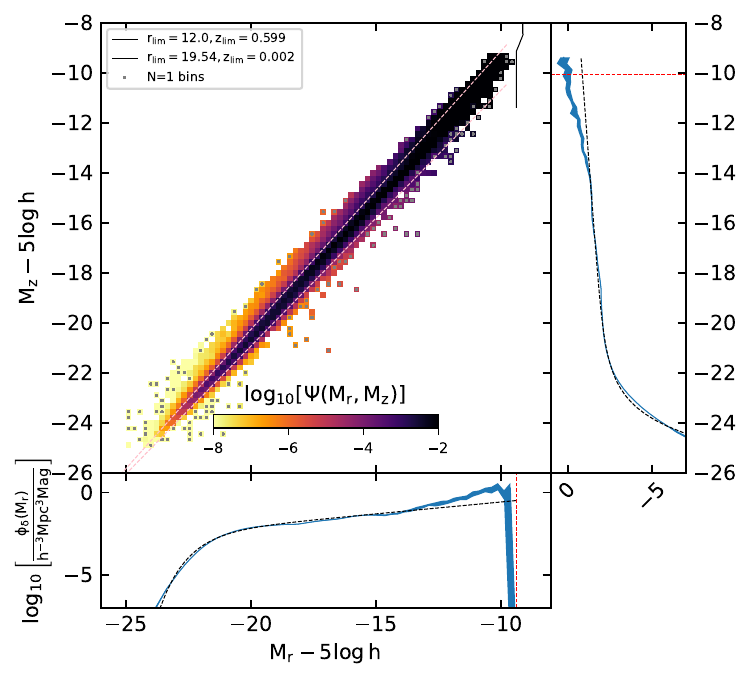}%
}
{%
  \label{fig:Num_Raw_scatter}%
  \includegraphics[width=0.46\textwidth]{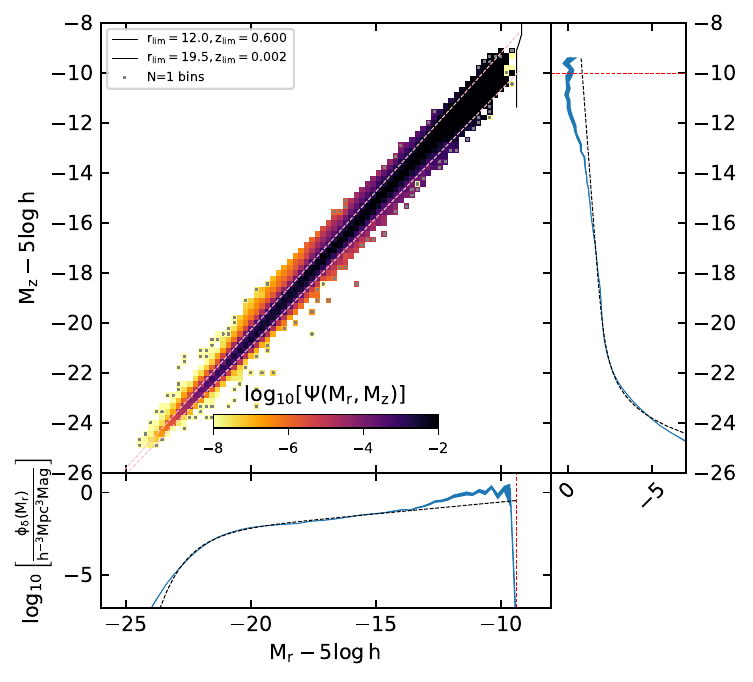}%
}%
{
  \label{fig:Total_Raw_scatter}%
  \includegraphics[width=0.46\textwidth]{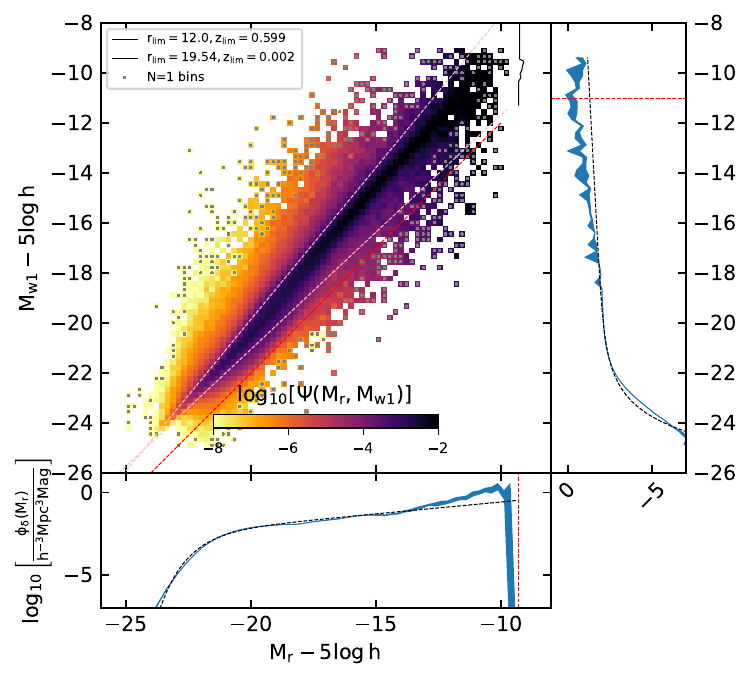}%
}%
{
  \label{fig:Num_Total_scatter}%
  \includegraphics[width=0.46\textwidth]{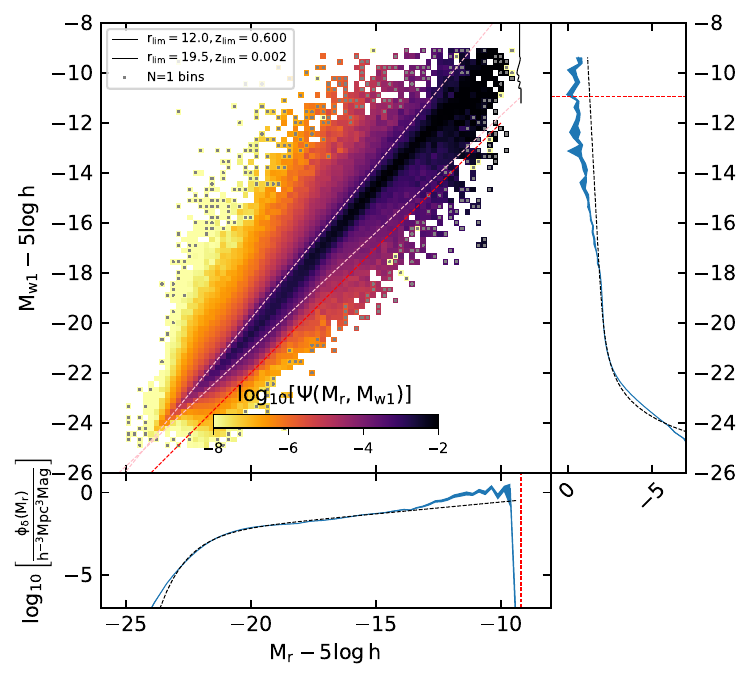}%
}
\caption{Bivariate LFs for North (left) and South (right) in the $g$, $z$ and $w1$ bands. The black curves represent the magnitude limit of the survey at the chosen $z=0.002$ limit. The dashed vertical and horizontal lines show the completeness limits for the $r$-band and $g$-band luminosity functions. These are based on the intersection of the completeness curves and the 95\% percentile contour of the bivariate LF. The $r-w1$ bivariate LFs have an additional dashed red line showing a selection cut that is later incorporated to remove spurious objects. Here, all LFs are presented with Poisson errors. The units of the LFs are $h^{3} \rm{Mpc}^{-3} \rm{mag}^{-1}$.}\label{fig:bivariate_LF}
\end{figure*}

With our methodology now described, we present our main findings. We generate a number of bivariate LFs in the $g$, $r$, $z$, and $w1$ bands (see Fig.~\ref{fig:bivariate_LF}) as well as the corresponding univariate LFs in those same bands (see Fig.~\ref{fig:global_LF}). We also present this LFs split by colour.
% deleted (see Fig.~\ref{fig:colour_LF})

\subsection{Bivariate LFs in \texorpdfstring{$g, r, z$ and $w1$}{g, r, z and w1}}

Our bivariate LFs are particularly useful for better understanding the completeness of the LFs that are presented in this paper. Noting that the BGS Bright survey is limited to $r=19.5$ in the South and $r=19.54$ in the North, we use the bivariate LFs to estimate the completeness of the $g$, $z$ and $w1$ LFs. Using the $g-r$ bivariate LF as an example, we do this by finding the $5\%$ to $95\%$ values of the distribution of $g$-band magnitudes in each $r$-band magnitude bin. Then, we plot the object-weighted least-squares fit regression lines to these percentiles, shown as the dashed pink lines in Fig.~\ref{fig:bivariate_LF}. 
As a black curve, we plot the absolute $r$-band magnitude limit at the chosen
minimum redshift limit ($z=0.002$) for a range of rest frame $^{0.1}(g-r)$ colours at the faint apparent magnitude limit of the sample
using Eqn. \ref{eq:mags}. 
The intersection of the limit with the 5\% regression line gives us an estimate of the completeness limit of the $g$-band LF, as shown by the horizontal line in the right-hand side panel. We follow the same process for the $z$ and $w1$ bands.

In addition, the bivariate LF offers a useful way to identify spurious objects. Bins with only a single object are marked by a red dot. As shown in Fig.~\ref{fig:bivariate_LF}, the $r-w1$ bivariate LF contains spurious peaks in the $w1$ LF. Our analysis shows that these peaks are caused by a small number of galaxies with a low $V_{\rm max}$ and a high value of $w_\textrm{comp}$ which in combination cause such galaxies to have a disproportionate effect on the LF. We visually inspect these galaxies using the Legacy Survey Sky Browser\footnote{\url{https://www.legacysurvey.org/viewer}}
and confirm that these galaxies are spurious, with fragmentation of large galaxies being a common issue. Typically,
each spurious peak in the LF corresponds to one such galaxy.
We note that these objects correspond to an unrealistic $\zref{(r-w1)}$ rest-frame colour. In order to deal with this, we add a colour selection cut which corresponds to \update{a rest-frame colour} selection cut of $\zref{(r-w1)} < 2.25$. This is shown as the dashed red line in the bottom two panels of Fig.~\ref{fig:bivariate_LF}. This selection cut is only applied to $w1$-band LFs.

\begin{table}
\centering
\caption{The Schechter parameters for the global $0.002<z<0.6$ LFs, used in Figs.~\ref{fig:global_LF} and \ref{fig:colour_LF} as fiducial reference curves. The meaning of the parameters are defined in Eqn.~\ref{eqn:single_sch}. We do not provide errors for these parameters as they are formally poor fits, as illustrated, for example, in Fig.~\ref{fig:LF_z_comp}.
}\label{tab:Sch}
 \begin{tabular}{||c c c c||} 
 \hline
  & $\log_{10} \Phi^*$ [$h^{-3} \textrm{Mpc}^3$] & $M^* - 5 \log h$ & $\alpha$\\ [0.5ex] 
 \hline\hline
 All (South) & & & \\ 
 %$g$ & -2.14 & -20.43 & -1.39 \\
$g$ & -1.91 & -19.98 & -1.26 \\
 $r$ & -2.06 & -20.97 & -1.28 \\ 
 $z$ & -2.15 & -21.88 & -1.28 \\
 $w1$ & -2.13 & -21.78 & -1.20 \\ [1ex]  
 \hline
 \end{tabular}
\end{table}

\subsection{LFs in  \texorpdfstring{$g, r, z$ and $w1$}{g, r, z and w1}}
\label{sec:LFs}

\begin{figure}
    \centering
    \includegraphics[width=0.95\columnwidth]{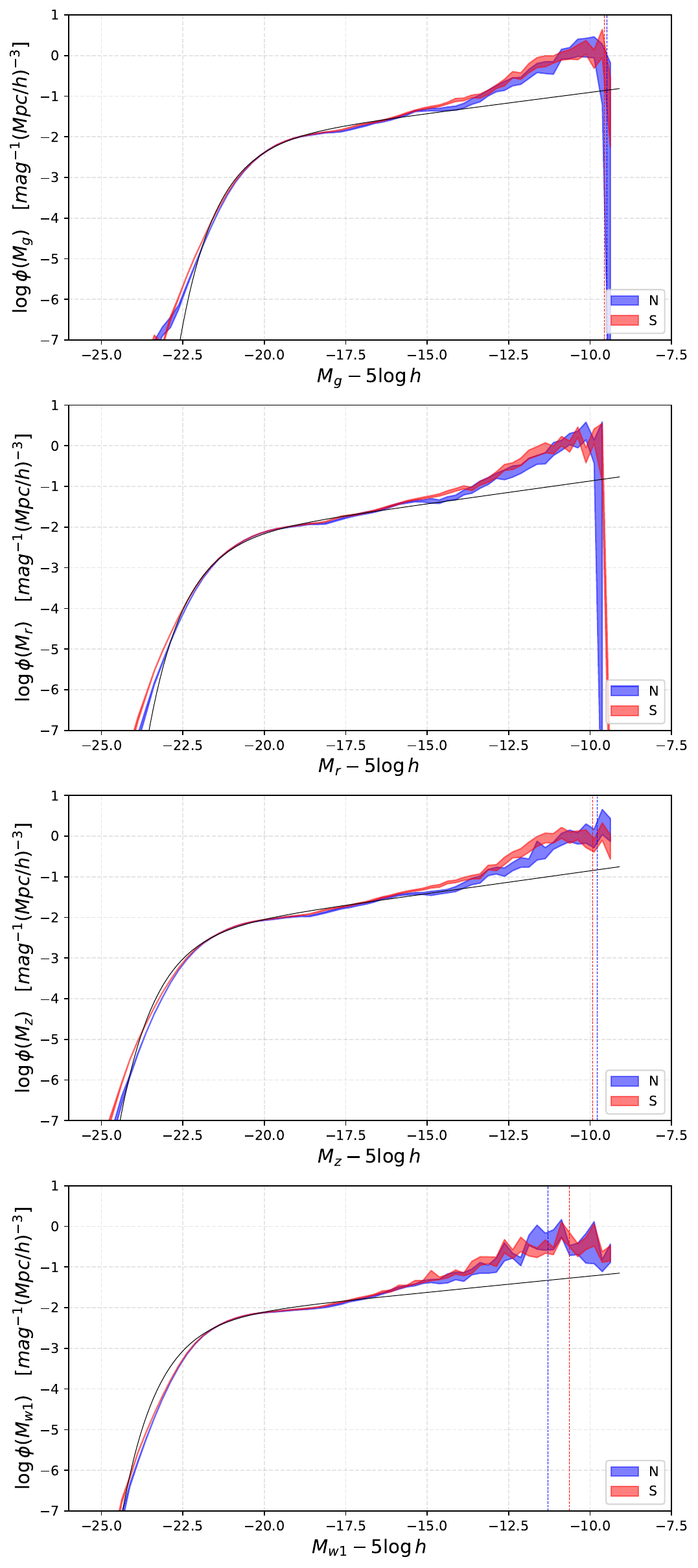}
    \caption{The global $1/V_\textrm{max}$ LF for Y3 data for North and South in the $g$, $r$, $z$ and $w1$ bands. The width of each LF represents the jackknife error. The dashed lines represent the completeness limits derived from the corresponding bivariate LFs. For reference, the black curves show simple Schechter function fits for each band.}
    \label{fig:global_LF}
\end{figure}

After generating the bivariate LFs shown in Fig.~\ref{fig:bivariate_LF}, we present the $1/{V_\textrm{max}}$ LFs in the $g$, $r$,  $z$, and $w1$ bands in Fig.~\ref{fig:global_LF}.  %\todo{In this figure and the next I don't think we should be plotting the estimate so far beyond where we think it is complete. We should at least stop before the point the LFs plummet.}
\footnote{The LFs in Fig.~\ref{fig:global_LF} differ from the projected LFs in Fig.~\ref{fig:bivariate_LF} in the following ways: 1) the Fig.~\ref{fig:global_LF} LFs have had the $\zref{(r-w1)} < 2.25$ selection cut applied as described above; 2) the thickness of the shaded bands in Fig.~\ref{fig:global_LF} represents the jackknife error in the LFs, whereas the thickness of the shaded bands in Fig.~\ref{fig:bivariate_LF} represents the Poisson error in the LFs.}

As the Y3 BGS survey has such a large number of galaxies, we note that the statistical errors on the LFs are extremely small.
As a consequence no simple functional form provides an adequate fit. This is illustrated by the Schechter functions we plot as fiducial curves in Fig.~\ref{fig:global_LF}. These fits were performed using the jackknife errors and by minimising $\chi^2$ over a limited magnitude range.
The parameters of these formally poor fits (e.g $\chi^2/\text{d.f.}=65.7$ over the full magnitude range for the South $r$-band LF)
are given in Table.~\ref{tab:Sch}.  
We strongly emphasise that we do not present these Schechter fits as a useful parametrisation of the LF.
While they are not good representations of the data, they are useful for highlighting systematic deviations from the Schechter function form and as fiducial references curves in subsequent plots. In terms of magnitudes the Schechter function form is
\begin{equation} \label{eqn:single_sch}
    \Phi(M) = \frac{\ln 10}{2.5} \Phi^* 10^{0.4(1+\alpha)(M-M^*)} \exp[-10^{0.4(M-M^*)}],
\end{equation}
where $M^*$ parameterises the position of the ‘knee’ of the Schechter function, $\alpha$ parameterises the faint end slope, and $\phi^*$ represents the space density around the ‘knee’.

We present LFs down to $^{0.1}M_r - 5 \log h \sim -10$, and go even fainter in the other bands. We note that this goes beyond that of comparable large scale surveys in the $r$-band, for example, SDSS reached a faintness of \update{ $-12$ \citep{2005ApJ...631..208B}
while the much smaller GAMA survey reached $-11$ \citep{Loveday2011}}. Over nearly all this range, our independent estimates from the North and South regions agree well with each other, indicating that the systematic dip below the Schechter function forms at a magnitude of approximately $-17$ is not caused by systematic errors.
However the faint upturn at $-14$, although seen in other work \citep{10.1093/mnras/stw898} is probably mainly the 
result of systematic errors or biases.

\begin{figure}
    \centering
    \includegraphics[width=\linewidth]{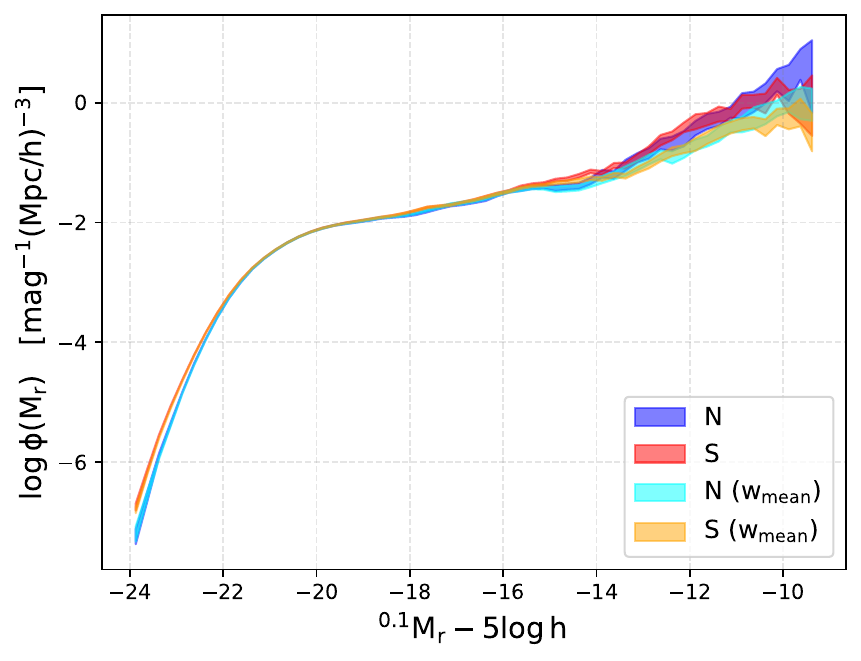}
    \caption{The global $1/V_{\rm max}$ $r$-band LFs for North and South. The LFs use either the standard weight given in Eqn.~\ref{eq:weights} where each galaxy has an individual total incompleteness weight, or the mean total weight ($w_{\rm mean}$) is assigned to all galaxies. The width of each LF represents the jackknife error.}
    \label{fig:weight_mean}
\end{figure}

\newupdate{
We investigate the impact of the targeting and redshift completeness weights on the shape of the LF by comparing our main results to an alternative LF in which every galaxy is assigned the mean incompleteness weight, $\bar{w} = 1.5275$. This comparison is shown in Fig.~\ref{fig:weight_mean}. For magnitudes brighter than $M_r \approx -16$ , the two LFs are indistinguishable, and the discrepancy between the North and South LFs at the bright end remains unchanged. This demonstrates that these features are not artefacts introduced by the redshift incompleteness weighting. At the faint end, however, the redshift incompleteness corrections are important. We find that the upturn at $M_r > -14$  becomes more pronounced when the individual weights are used. This behaviour arises because intrinsically faint galaxies tend to have lower surface brightness and fainter fibre magnitudes, placing them in the region (see Fig.~\ref{fig:weight_lookup}) where the incompleteness corrections are largest.
}

\newupdate{We also investigated whether the limiting surface brightness of the Legacy Survey photometric data and the explicit fibre magnitude limit in the BGS selection
causes our luminosity functions to be incomplete. In the DESI Large Scale Structure analysis, fake galaxy images have been superimposed on the imaging data to gauge the ability of the photometric pipeline to recover them as a function of position on the sky \citep[e.g. see][]{2025JCAP...01..146K}. This analysis did not explicitly investigate low surface brightness galaxies. It would be valuable to undertake such an investigation. However,  while such an approach can accurately quantify the limits of the data, it cannot directly tell us whether there is a population of galaxies that we are missing. We have therefore used a region of much deeper data from the COSMOS deep drilling fields to directly estimate how many galaxies we
are missing. This analysis is presented in Appendix~\ref{appendixSB}, together with an assessment of the impact of the BGS fibre magnitude limit.

Fainter than $M_r-5 \log h \approx -15$, we find a small tail of low-surface brightness objects in the deeper data whose inclusion would boost the faint end of the LF by between 0.5 and 1-$\sigma$. However, inspection of these images in the Legacy Survey data reveals that about half of them are associated with diffraction spikes or lie within the images of larger galaxies.
Consequently, we have chosen not to apply this uncertain correction to our estimates. Separately, based on a simple parametric model of the galaxy surface brightness distribution we find that the fibre-magnitude cut causes a systematic underestimation of the LF which becomes comparable but not larger than the statistical errors fainter than  $M_r-5 \log h \approx -14$.
We of course cannot rule out a unseen population of much lower surface brightness galaxies that are not detectable in the deeper data or captured in the assumed Gaussian distribution of our simple parametric model.
}

In Appendix~\ref{appendixLFmethods}, we present the Stepwise Maximum Likelihood (SWML) and $1/V_{\rm dc, max}$ LFs alongside the $1/V_{\rm max}$ LFs (see Fig.~\ref{fig:LF_method}). We find that the upturn is present in the LF in all three of the methods, but it is substantially reduced when using the SWML and $1/V_{\rm dc, max}$ estimators - which are designed to correct for density fluctuations in the LF. This suggests that most of the faint-end upturn is due to local density fluctuations. 

\update{We also investigated if some of the upturn is caused by fragmented galaxy images or other image analysis problems. In Appendix~\ref{tableVIimaging} we present the details of a visual inspection programme. We find that some of the faintest galaxies are likely to invalid targets, often features within larger galaxies. However the effect on the LF is modest, restricted to the faint end and does not remove the steepening of the faint end of the LF that we have reported.    }

%We conducted further visual inspection of the faint galaxies using the Legacy Survey Sky Browser. We find that there are a significant number of spurious galaxies. For example, when inspecting all 41 galaxies with $^{0.1}M_r - 5\log h > -10$ we find 9 are likely to be galaxies (even if the magnitudecould be inaccurate), 3 that are possibly galaxies, and 29 are clearly spurious. \update{The full results of this analysis are shown in Appendix~\ref{tableVIimaging}}. In particular, thevast majority of problematic galaxies appear to exist within another more dominantgalaxy or dust cloud. As such, we raise the concern that many of the objects at thefaintest magnitudes may be artificial and erroneous. For example, there is the issueof fragmentation, where a single extended source is targeted multiple times. As aresult of this, there may exist multiple targets for a single large galaxy where each target gets a lower flux than expected. By randomly sampling galaxies in each magnitude bin of the LF and visually inspecting them, we believe that a more conservative limit to the $r$-band LF is around $-12$, as this is where we start to observe more likely galaxies than spurious galaxies in our inspections. At fainter magnitudes, there exist a large number of bins that are dominated by problematic objects. We carefully note that this leads to the possibility that the upturn in the global LF is a consequence of the imaging analysis artefacts. Specifically, fragmentation leads to a larger number of objects being counted in the fainter magnitude bins. }

We also attempted to use a Convolutional Neural Network to classify and remove these spurious objects. We found that doing so flattened the upturn, but did not completely remove it. As such, it is still possible that the upturn is a real feature in the global LF, but we recognise that further investigation is required to confirm this.

 The most comprehensive set of LF estimates previously was that of the GAMA survey \citep{Loveday2011}. We have not plotted the
 Loveday results on Fig.~\ref{fig:global_LF} as with such large observational samples the systematic differences coming from differing choices of redshift range and evolutionary assumptions dominate over the statistical differences. 
 In Appendix~\ref{appendixLovedaymethods} we demonstrate
 that when making the same assumptions we find our estimates are
 excellent agreement with those of \cite{Loveday2011}
 albeit with much smaller statistical errors. 
  Also in 
 Appendix~\ref{appendixLovedaymethods}, 
 we show that 
  splitting the sample by redshift reveals small but highly significant residuals that indicate our
simple global evolution model, while being accurate around the knee of the LF, fails to capture the more complex evolution that
the BGS reveals.

 %In 
 %Appendix~\ref{appendixLovedaymethods} we also investigate
 %the sensitivity of our estimate to our adopted model of %evolution.
 %This indicates that our e-correction model may be in need of improvement. We find that a two-class e-correction model using a different $Q$ for red and blue galaxies results in a negligible improvement.

\begin{figure}
    \centering
    \includegraphics[width=\columnwidth]{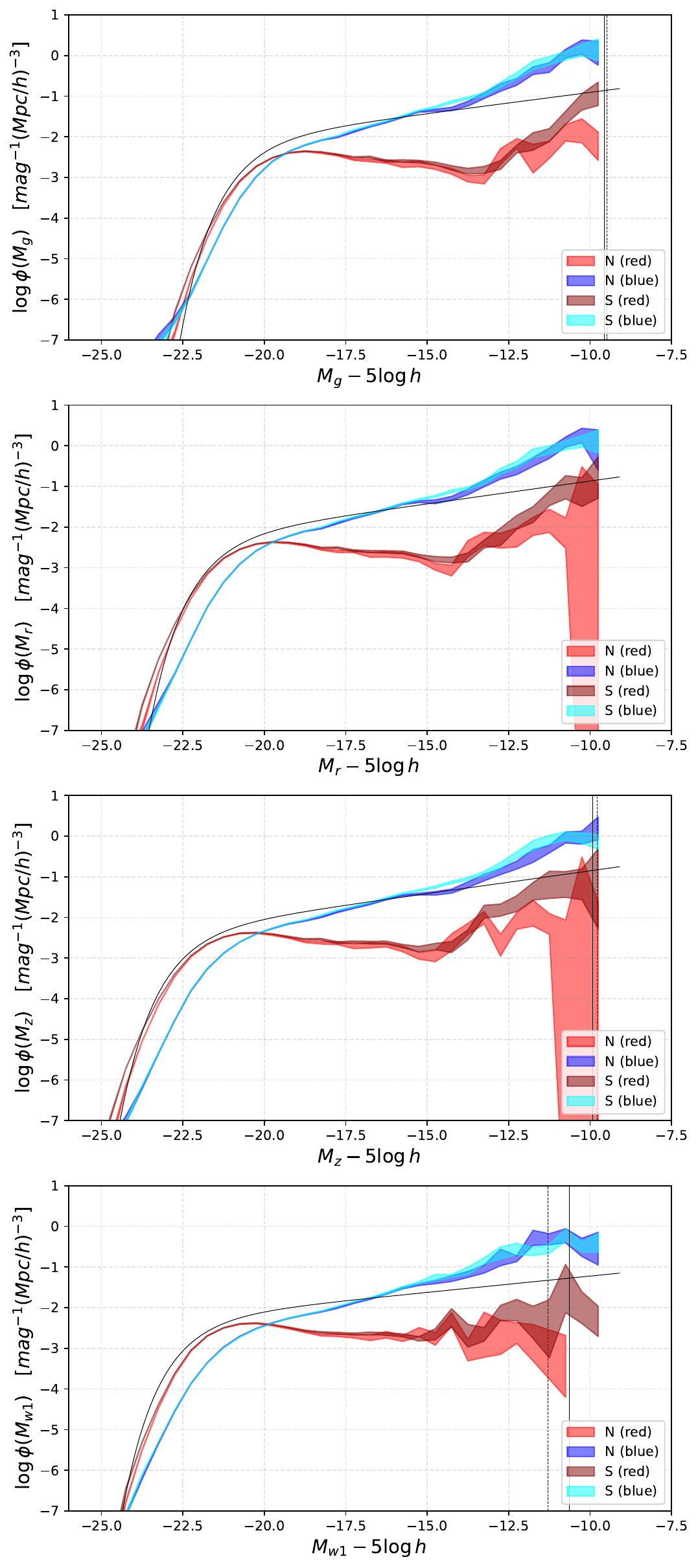}
    \caption{The global $1/V_\textrm{max}$ LF for Y3 data for North and South in the $g$, $r$, $z$ and $w1$ bands split by colour. The red LFs represent galaxies with $\zref{(g-r)} > 0.75$, while the blue LFs are galaxies with $\zref{(g-r)} < 0.75$. The width of each LF represents the jackknife error. The dashed lines represent the completeness limits derived from the corresponding bivariate LFs. The solid fiducial Schechter curves are reproduced from Fig.~\ref{fig:global_LF}. Their parameters are given in Table~\ref{tab:Sch}.}
    \label{fig:colour_LF}
\end{figure}

While over most of the magnitude range, we observe that there exists excellent agreement between the North and South LFs we note that there exists a systematic discrepancy between the North and South $g$, $r$ and $z$ LFs at the bright-end of the LFs (e.g: $^{0.1}M_r - 5\log h < -22.5$). We extensively investigated the cause of this highly statistically significant difference.
We first investigated the impact of the previously noted 
rest-frame colour discrepancy between North and South (as seen in Fig.~\ref{fig:col_mag}, noting that this distributional difference is consistent with that seen in the FSF catalogue and is not caused by our k-correction methodology). To force the colour distributions to be the same we made an empirical correction by shifting the $r$-band magnitudes of the galaxies in the North such that we preserve their ranking in rest-frame colour and then exactly match the cumulative rest-frame colour distribution of the South. This made a negligible difference to the LFs, showing that whatever is perturbing the rest-frame colour distributions is not responsible for the LF discrepancy. Instead, we believe that this is at least partially caused by a difference in the light profile fitting between North and South. 
We investigate the photometric differences between North and South in Appendix~\ref{appendixFSF}. 

In the deep DECaLS photometry of the South
some very luminous early-type galaxies are fit with light profiles
with  extreme Sérsic profiles ($n_\textrm{sérsic}>4$) and hence very extended outer light profiles. Extended wings are not detected in the shallower BASS/MzLS data and so are fitted with less extreme Sérsic profiles and result in fainter magnitudes. Whether this represents a limit on of the shallow North data or issue of the inadequate sky subtraction in the more demanding lower noise South data is unclear.
Hence we conclude that this North-South difference represents a not yet fully understood systematic uncertainty in the bright end of the galaxy LFs.

\subsection{LFs split by  \texorpdfstring{$\zref(g-r)$}{g-r} colour}

In Fig.~\ref{fig:colour_LF}, we present the global luminosity functions (LFs) split by rest frame colour. To obtain colour-dependent LFs, we separate our sample into red and blue populations using the rest-frame colour-cut $\zref{(g-r)} = 0.75$, as defined in Section~\ref{e-corrections}. 
We show red and blue $1/V_\textrm{max}$ LFs for each band
and note that the red and blue populations are consistently defined by $\zref{(g-r)} = 0.75$, even when examining the $z$ and $w1$ bands.
 For the colour-dependent LFs we show completeness limits that correspond to the global completeness limits found in Fig.~\ref{fig:bivariate_LF}. We assume that these completeness limits are conservative enough to be applicable to the colour-split LFs. 

 We observe that the bright-end offset in the absolute magnitudes is predominantly seen in the red population, with a visible difference between the red N and S LFs. The corresponding difference for the blue N and S LFs is minimal.
This is consistent with the hypothesis that for the brightest galaxies
in the shallower N data, the contribution of the total magnitude of the outer profile is underestimated. For blue galaxies which typically have exponential profiles, very little flux is contributed from large radii. However, for bright red galaxies, which are typically ellipticals with either de Vaucoleur profiles or high-$n$ Sérsic profiles, a significant fraction of their light is contributed from their outer profiles. Moreover, as a single profile is fitted across all of the Legacy Survey bands, we expect and see the same offset in the $g$, $r$ and $z$ bands. We do not observe an offset between N and S in the $w1$-band. This is explained by the fact that the $w1$ PSF is much wider than in the other bands. As such, the PSF convolved profiles do not differ significantly for the typical bright distant galaxy.

 The colour-split LFs quantify well-known differences in the demographics of red and blue galaxies. At around magnitude $-20$ red and blue galaxies are equally abundant but their luminosity functions are quite different with red galaxies dominating at bright magnitudes and blue galaxies at faint magnitudes. This is true in all bands. The turn up in the luminosity functions at faint magnitudes is most dramatic for red galaxies with the slope of the luminosity function changing sign at around a magnitude of $-13$. Despite the very steep faint end of the red-galaxy luminosity function this population is still sub-dominant at the faintest magnitudes we probe. Hence
 the faint-end turn up in the overall luminosity function is dominated by a corresponding turn up in the faint-end slope of the blue-galaxy luminosity function.

\section{Conclusions}
\label{Conclusions}

In this paper, we have presented the $g$, $r$, $z$ and $w1$ LFs using the DESI Y3 data. The large number of galaxies in the Y3 BGS dataset mean that we have constrained the errors on the LFs further than prior studies in the literature. Furthermore, these LFs provide useful measurements to very faint absolute magnitudes, extending the LF to extremely faint magnitudes (such as $^{0.1}M_r - 5\log h  \sim -10$). 
\newupdate{Although an assessment of incompleteness for low surface brightness galaxies indicates that the LF estimate from the BGS is only complete for galaxies with $r$-band half-light surface brightness brighter than $\mu_{50}>25$ and that fainter than  $^{0.1}M_r - 5\log h  \sim -14$,
this impacts the LF estimate at a level comparable to the reported statistical errors.
}

In order to produce these LFs, we have developed a robust methodology for generating k-corrections for galaxies in order to calculate absolute magnitudes. In addition, we incorporate a e-correction model that uses a single value of $Q$ for all galaxies. We broadly find that this is sufficient enough for our purposes, but note that this does not fully incorporate the entire evolutionary process, as shown by a discrepancy between our $0.002<z<0.1$ and $0.002<z<0.6$ $r$-band LFs in Fig.~\ref{fig:LF_z_comp}.
%Whilst experimenting with a two-value $Q$ model (using $Q_{\rm red}$ and $Q_{\rm blue}$) saw marginally improved results, it was not sufficient to fully incorporate the full evolution. 
As such, we acknowledge that further work is required to improve this aspect of our model but also note this illustrates the power of the DESI data to probe galaxy evolution.

We verify that our LFs yield similar results when different estimators are used. In particular, we find that the $1/V_{\rm max}$ LFs yield similar results to the SWML LFs and $1/V_{\rm dc, max}$ LFs, except at the very faintest absolute magnitudes. Overall, this methodology provides a robust platform for us to further constrain the LFs in the future Y5 DESI data releases.

We find that there are a number of galaxies with spurious properties in the Y3 sample. To begin, we find that there are a number of galaxies with abnormal $^{0.1}(r-w1)$ rest-frame colours. Through visual inspection, this was indicative of a number of bad objects which could be removed with a colour cut. However, further visual inspection reveals that there are a significant number of bad objects at faint $r$-band absolute magnitudes. This suggests that further work is needed to understand the imaging at the faintest magnitudes. In addition, this acts as an additional constraint on the completeness of our luminosity functions. For example, while our bivariate LFs suggest that our $g$-band LFs are complete up to fainter than $^{0.1}M_g - 5\log h= -10$, we suggest that this is a very liberal estimate, and accounting for poor imaging will bring the completeness limit down.

As part of our investigation, we found that there exists a visible disparity between the North and South LFs at the bright-end when using the DESI total magnitudes. When adjusting our methodology to find Petrosian magnitude LFs, we were able to \update{somewhat reduce} the bright-end disparity between the LFs.  We believe that this difference is explained by the shallower North photometry being unable to properly measure the faint outskirts of objects. As the Petrosian magnitudes account for the Sérsic index, this explains why they are successful in reducing the bright-end disparity.

The small statistical errors of our estimates result in the LFs not being well fit by simple analytic forms.
They exhibit a bright end that is not well fit by an exponential and complex non-powerlaw behaviour faintward of the knee. In particular, we observe the existence of an upturn in the LFs at 
$^{0.1}M_r - 5 \log h \ge -15$, which is strongest for red galaxies.

\section*{Acknowledgements} \label{Acknowledgements}

 This material is based upon work supported by the U.S. Department of Energy (DOE), Office of Science, Office of High-Energy Physics, under Contract No. DE–AC02–05CH11231, and by the National Energy Research Scientific Computing Center, a DOE Office of Science User Facility under the same contract. Additional support for DESI was provided by the U.S. National Science Foundation (NSF), Division of Astronomical Sciences under Contract No. AST-0950945 to the NSF’s National Optical-Infrared Astronomy Research Laboratory; the Science and Technology Facilities Council of the United Kingdom; the Gordon and Betty Moore Foundation; the Heising-Simons Foundation; the French Alternative Energies and Atomic Energy Commission (CEA); the National Council of Humanities, Science and Technology of Mexico (CONAHCYT); the Ministry of Science, Innovation and Universities of Spain (MICIU/AEI/10.13039/501100011033), and by the DESI Member Institutions: https://www.desi.lbl.gov/collaborating-institutions. Any opinions, findings, and conclusions or recommendations expressed in this material are those of the author(s) and do not necessarily reflect the views of the U.S. National Science Foundation, the U.S. Department of Energy, or any of the listed funding agencies.

The authors are honored to be permitted to conduct scientific research on I'oligam Du'ag (Kitt Peak), a mountain with particular significance to the Tohono O’odham Nation.

SC, PN and MW acknowledge the support of STFC consolidated grant ST/X001075/1.
SM acknowledges the support of STFC  grant ST/V506643/1.

\update{We thank Hui Kong for useful advice and technical help with supplementary datasets.}

\section*{Data access statement}

All data, unless explicitly stated in text, is sourced from the DESI collaboration. The data used in this analysis will be made public along the Data Release 1 release (details found at https://data.desi.lbl.gov/doc/releases/). Documentation of DESI data access is maintained at https://data.desi.lbl.gov/doc/access/. The data points for our LFs in Fig.~\ref{fig:global_LF}, Fig.~\ref{fig:colour_LF},
Fig.~\ref{fig:LF_method} and Fig.~\ref{fig:LFgzw1_SWML}
are accessible at https://icc.dur.ac.uk/data/.

\appendix
\section{Low Surface Brightness Incompleteness} \label{appendixSB}

The median 5-$\sigma$ $r$-band depth for a fiducial galaxy in the Legacy Survey DR9 data release, that was used to select DESI targets, is 23.9~mag in the South (DECaLS) and  23.5~mag in the North (BASS) \citep{Dey_2019}.  While this is much 
deeper than 19.5 magnitude limit of the BGS-BRIGHT sample we are analyzing, it is possible that the catalogue is incomplete for low surface brightness galaxies. To assess this we have made use of the much deeper COSMOS deep drilling fields \citep{2023MNRAS.519.3881G}. These data are comprised of hundreds of additional DECam images that have been analysed with the same Legacy Survey pipeline \citep{Dey_2019} to produce much deeper catalogues over tens of square degrees centred on the COSMOS field \citep{2007ApJS..172....1S}. In this analysis we extract the approximately 20 square degrees that has a depth greater than 24.5~mag.

\begin{figure}
\includegraphics[width=\columnwidth]{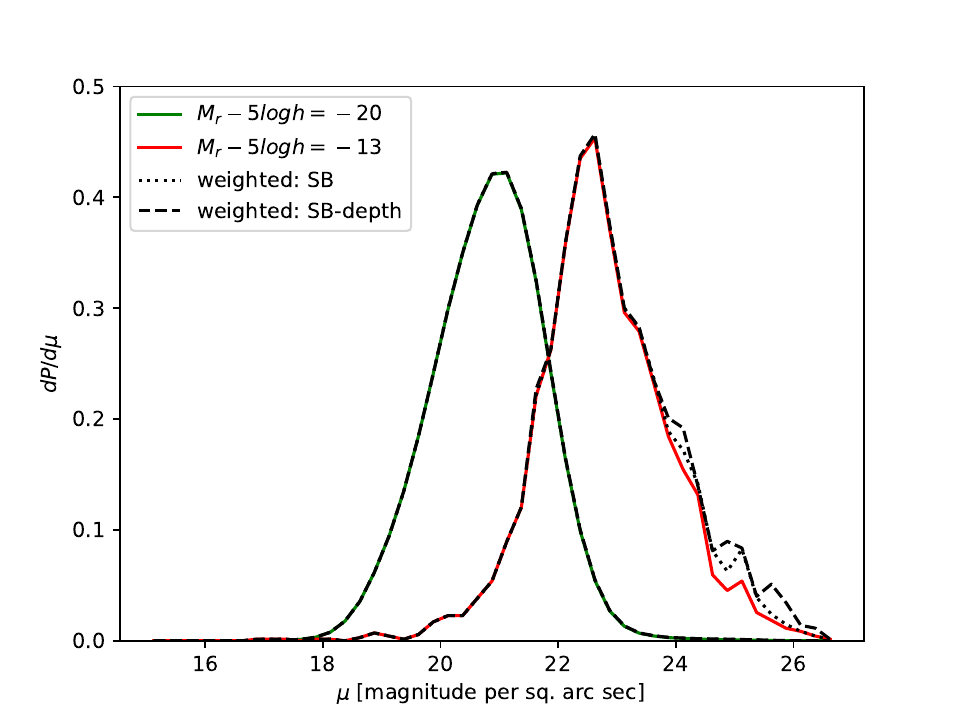}
    \caption{The solid curves show the normalised distributions of half-light surface brightnesses for galaxies from Legacy Survey DR9 (the parent catalogue of the BGS) of absolute magnitude $M_r - 5 \log h=-20$ (green) and $-13$ (red).  The dotted and dashed lines show the effect on these distributions (not normalised) of applying the two up-weighting schemes to correct for incompleteness based on estimates from the deeper COSMOS data.}
        \label{fig:SBdist}
\end{figure}

\newupdate{
For all the galaxies in the Legacy Survery DR9 catalogue and sources in the deeper COSMOS data we have computed their
mean surface brightness within their elliptical half-light radii (without any correction for MW extinction).
The solid curves in Fig.~\ref{fig:SBdist} show the normalized surface brightness distributions of Legacy Survey galaxies 
selected in absolute magnitude bins centred at $-20$ and $-13$. 
In this comparison, we do not apply the explicit fibre-magnitude limit which is part of the BGS target selection \citep{Omar2020,Hahn_2023} as we wish to isolate only the effect of the photometric depth.
The distribution for the fainter galaxies peaks lower surface brightness and extends 26 mag per square arcsec. Thus if BGS is incomplete at low surface brightness it will preferential affect the faint end of the galaxy luminosity function.
}

To assess the incompleteness in BGS we took all non-PSF sources brighter than $r=20$ in the COSMOS data and spatially matched them using a  1.5~arc~sec search radius to the full DR9 Legacy Survey catalogue from which BGS was selected. The magnitudes of the matching objects agree well with a 1-$\sigma$ (half the 16 to 84\% width of the distribution)  of 0.13 mag even at magnitudes fainter than the $r=19.5$ limit of BGS. 
We matched to the full Legacy Survey rather than the BGS catalogue as otherwise COSMOS objects close to  $r=19.5$ would be missed in BGS just because of the random scatter in the difference of the two magnitude measurements and the cut at 19.5 imposed in BGS. As a measure of the completeness we tabulated the fraction of matched COSMOS objects as a function of apparent magnitude and surface brightness. The inverse of this can then be used as weight to apply to all BGS galaxies as a function of their apparent magnitude and surface brightness. 
As the depth of the Legacy Survey has some variation across the sky we also tabulated the completeness as a function of apparent magnitude and surface brightness minus depth. We use this to define an alternative weight to correct for the incompleteness while taking some account of the depth variations.

The dashed and dotted lines in  Fig.~\ref{fig:SBdist} show the surface brightness distributions after applying these incompleteness corrections. One can see  that this makes negligible difference to the distribution at 
$M_r - 5 \log h = -20$ as essentially no objects with surface brightness brighter than 24~mag per square arcsec are missed by the Legacy Survey.  For $M_r - 5 \log h = -13$, one can see that the faint side of the distribution is modestly boosted, especially for the depth-dependent weight.
Fig.~\ref{fig:SBcorrectedLF} shows the effect of applying the depth dependent weight on the resulting $r$-band LF.
As expected this correction only affects the faint end of the galaxy luminosity function where it causes a systematic increase in the LF at the level of the 1-$\sigma$ statistical error. 
While this is not negligible we have chosen to not apply this correction as when we view the Legacy Survey images of the missed objects a significant fraction of them appear to be spurious artefacts. We found 20\% were associated with diffraction spikes and 30\% were within the image of an already detected larger galaxy (see Manwadkar et al. (in prep) for detailed study of the shredding of larger BGS galaxies) .

\newupdate{
Another important factor affecting the representation of low surface brightness galaxies in the BGS catalogue is the explicit fibre-magnitude cut that is part of the BGS target selection. Only objects with fibre-magnitude
\begin{equation}
r_{\mathrm{fib}} =
\begin{cases}
22.9 + (r - 17.8), & r < 17.8, \\
22.9,              & r \ge 17.8 .
\end{cases}
\end{equation}
\citep{Omar2020,Hahn_2023}
are included in the BGS.

\begin{figure*}
\includegraphics[width=18cm]{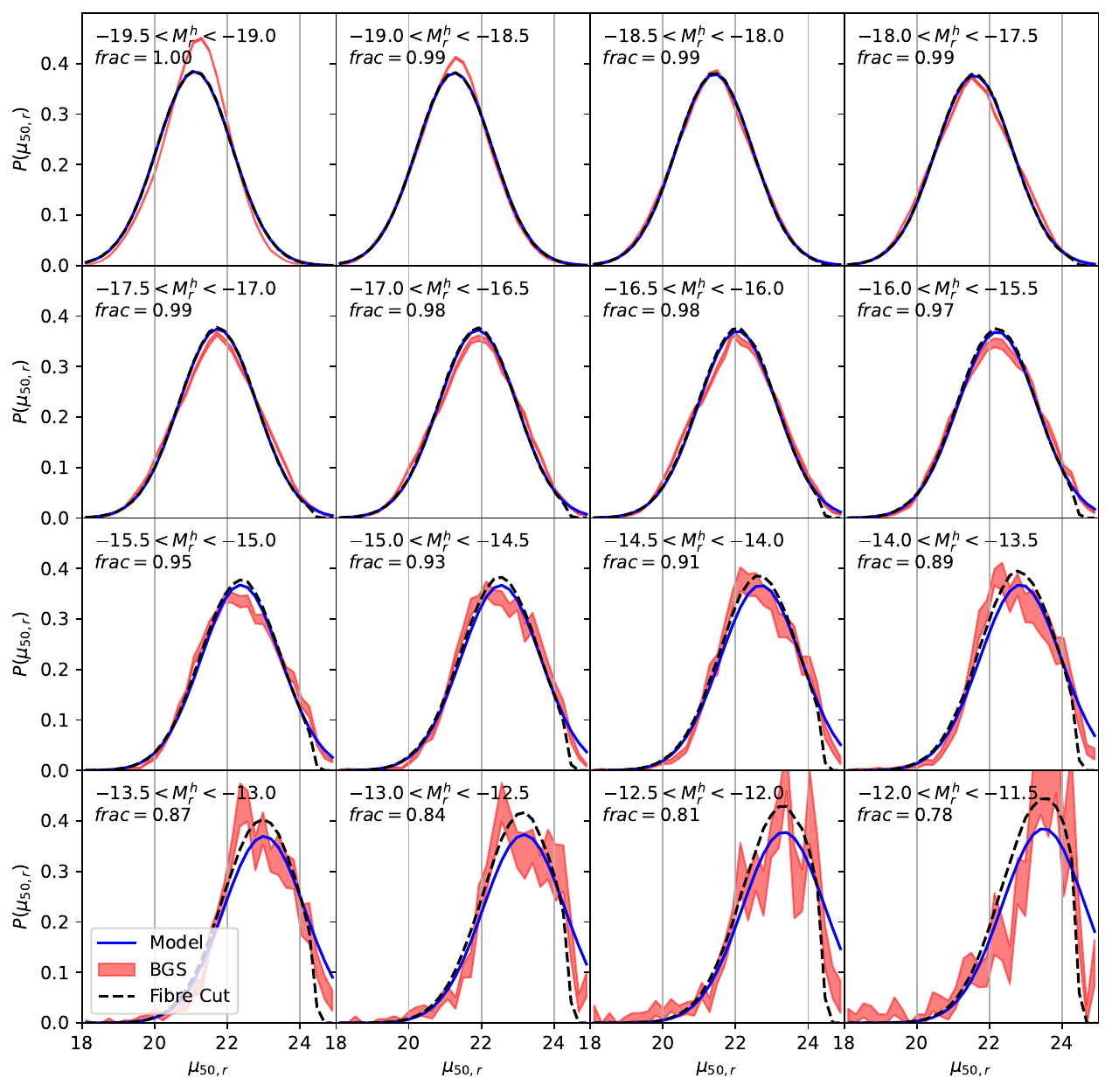}
    \caption{The $r$-band surface brightness distribution of BGS galaxies in bins of absolute $r$-band magnitude. The red bands show the redshift incompletness correct estimates from BGS with jackknife errors. The blue curves are a Gaussian model fit to the data in the range $-19<M_r-5 \log h <-15$ in which the mean and standard deviation of the Gaussian is assumed to vary linearly with magnitude. The black dashed curves show the affect on these distributions of applying the BGS fibre-magnitude cut. The fraction of galaxies satisfying this selection cut is shown on each panel.}
        \label{fig:SBdist_16}
\end{figure*}

The only way to estimate the incompleteness beyond this limit is with respect to an assumed model of the intrinsic surface brightness distribution. Here we construct such a model following  the procedure in  \cite{2005ApJ...631..208B}. In Fig.~\ref{fig:SBdist_16}, we show the $r$-band surface brightness distributions of the BGS galaxies with $n_textrm{Sersic}<2$ in bins of absolute magnitude. The surface brightness measure we use, $\mu_{50}$, is the mean surface brightness within the ellipse containing 50\% of the light. Over the range $-19<M^h_r<-15$ (where $M^h_r=m_r- 5 \log h$) we fit these data with a Gaussian distribution whose mean and standard deviation are linear functions of magnitude. The best fitting relation is
\begin{equation}
   \bar{\mu}_{50} = 20.703     + 0.319 (M^h_r +20.5) 
\end{equation}

\begin{equation}
\sigma_\mu = 1.022 - 0.014 (M^h_r+20.5)  .
\end{equation}
The blue curves in Fig.~\ref{fig:SBdist_16} show this model including its extrapolation to brighter and fainter bins.  The selection of only 
$n_\textrm{Sersic}<2$ galaxies follows   \cite{2005ApJ...631..208B} and is motivated by the goal of modelling the faint galaxy regime
where a large majority of the galaxies are have exponential profiles.

To estimate the incompleteness  in each absolute magnitude bin caused by the fibre-magnitude limit, we sample from the corresponding model distribution of surface brightness, assign each sampled object an $r$-band apparent magnitude, $r$, chosen at random from the corresponding BGS data. This ensures our model sample has the same distance distribution as the BGS data.
We then define effective radius $r_{50}$ for each object via the identity
$$
\mu_{50} = r +2.5 \log_{10} ( 2\pi r_{50}^2).
$$
To compute a fibre magnitude we assume each object has an $n_{Sersic}=1$ Sérsic profile, convolve with the fiducial 1~arcsec FWHM seeing and compute the 
magnitude in the 1.5~arcsec diameter fibre aperture.  This an appropriate choice of Sérsic profile for faint galaxies as the majority of such galaxies have profiles that are well fit by exponential profiles.
The distributions of surface brightness of the objects that satisfy the BGS fibre-magnitude cut are shown by the black dashed lines.

The fraction in the label on each panel is the fraction of the model galaxies satisfying the selection criteria.  While the agreement between the data and this
simple
model is not perfect,  we see no  evidence that there are fewer low surface brightness objects in the BGS data than predicted by this model. Hence the incompleteness fractions shown in each panel are reasonable estimates of the incompleteness caused by the fibre-magnitude limit. 

Faintward of $M_r-5 \log h = -14$,
the level of incompleteness indicated by this model becomes comparable, but does not exceed, the statistical errors of our estimated LFs. The incompleteness revealed by the deeper COSMOS deep drilling data should not be considered as addition to this as the model galaxies deselected by the fibre-magnitude cut could represent galaxies detected by COSMOS.

In summary, this investigation revealed only a modest level of incompleteness due to the depth of the Legacy Survey imaging and fibre-magnitude cut. However, the latter estimate is model dependent and so does not preclude an additional population of much lower surface brightness objects that could only be detected in data deeper than that of the COSMOS deep drilling fields. Our estimate should be taken as the galaxy luminosity function for galaxies with half-light surface brightnesses greater than approximately 25 magnitudes per square arcsec.
}

\begin{figure}
\includegraphics[width=\columnwidth]{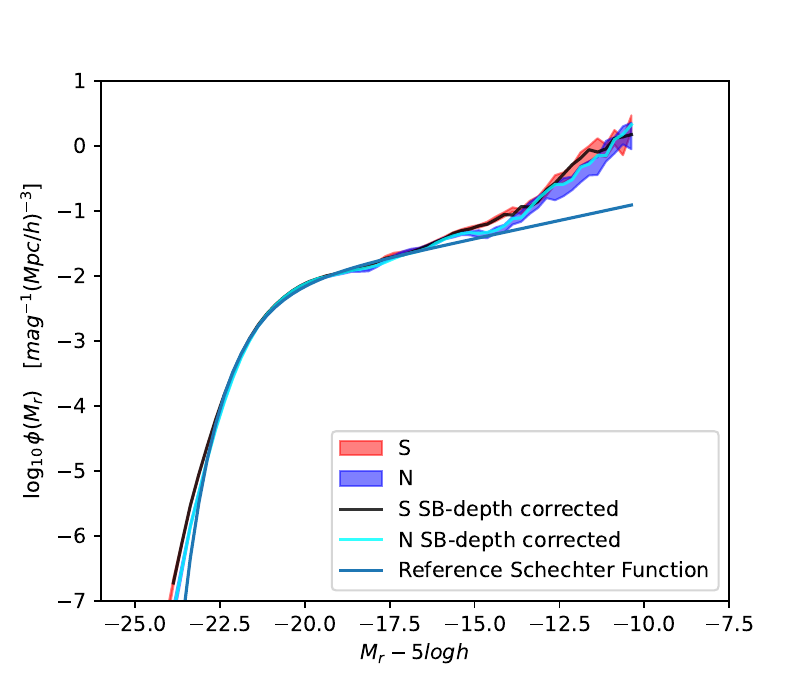}
    \caption{A comparison the $1/V_{\rm max}$ $r$-band LF before and after correcting for low surface brightness incompleteness. The red and blue shaded regions show the standard/uncorrected 1-$\sigma$ error interval for the South and North data respectively. The black and cyan lines show the corresponding estimates after applying the correction estimated from the COSMOS deep drilling fields. The correction factor used here was tabulated as a function of apparent magnitude and surface brightness minus survey depth.   }
        \label{fig:SBcorrectedLF}
\end{figure}

\section{Comparison of LF Methods} \label{appendixLFmethods}

To verify the robustness of the $1/V_{\rm max}$ LF method described in Section \ref{Luminosity Functions}, we compare to two other LF estimators. Specifically, we compare with the Stepwise Maximum Likelihood (SWML) estimator of \cite{1988MNRAS.232..431E} and a density-corrected $V_{\rm max}$ method ($V_\textrm{dc,max}$) estimator, which is based on a method from \cite{2011MNRAS.416..739C} described in more detail below. From this, we can confirm that brighter than $-14$, our LFs are largely invariant of the estimator used. Fainter than $-14$, the turn up of the LF is significantly reduced using the SWML and $V_\textrm{dc,max}$ estimators.

\begin{figure}
    \centering
    \includegraphics[width=\columnwidth]{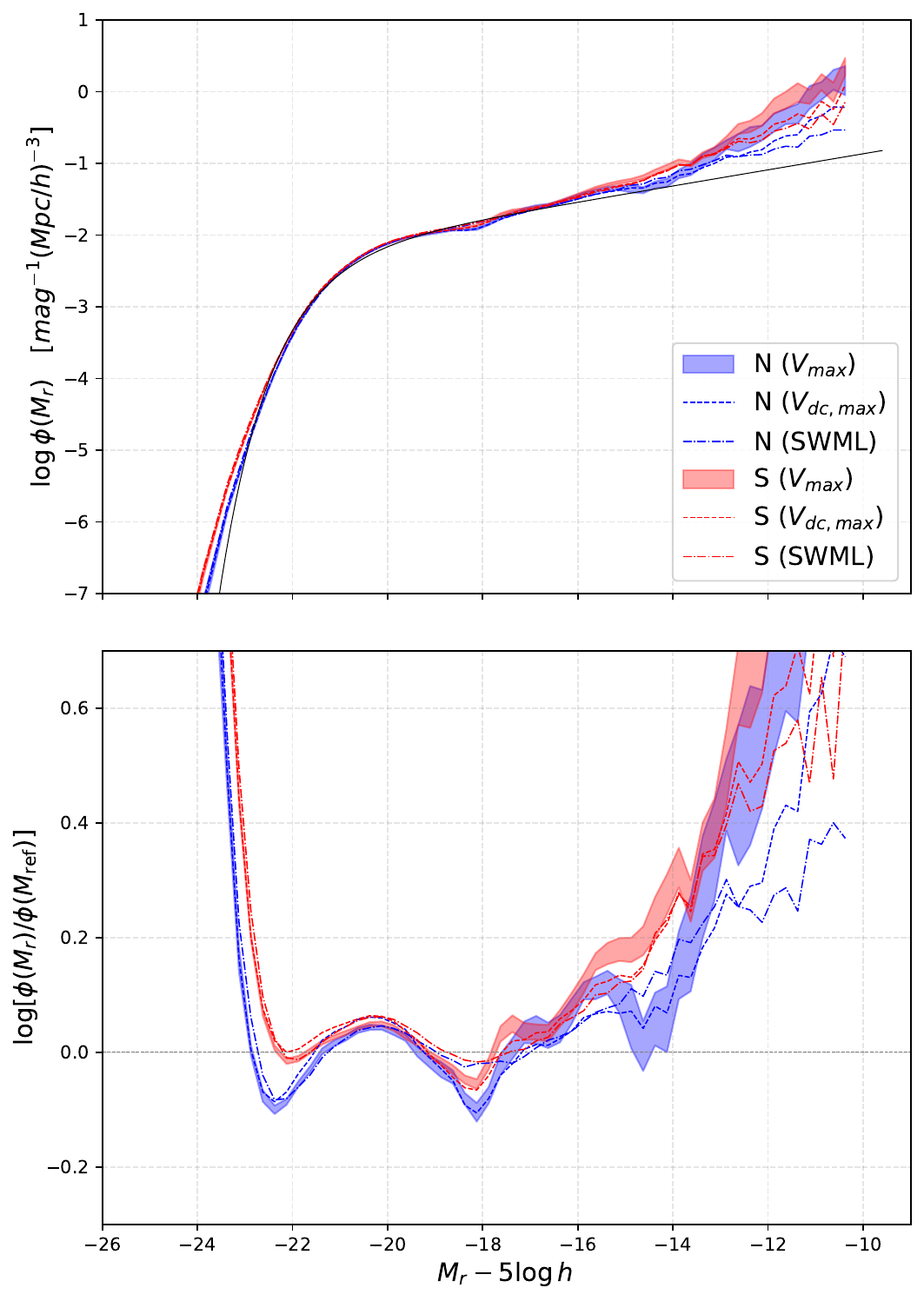}
    \caption{Top: A comparison of the different LF estimators ($1/V_{\rm max}$, $1/V_{\rm dc, max}$, SWML) in the North (blue)  and South (red). The black curve shows the $r$-band fiducial Schechter function from Table~\ref{tab:Sch}.
    Bottom: The ratio of each of the LFs estimates to fiducial Schechter function. }
    \label{fig:LF_method}
\end{figure}

 The SWML method does not require the assumption of a functional form of the LF (compared to other methods such as the Sandage, Tammann \& Yahil (STY) estimator). Moreover, the SWML estimator is unbiased by density fluctuations if one assumes that the LF shape is independent of density (unlike the $V_\textrm{max}$ method, which can be biased by density fluctuations). In our second paper Moore et al. in prep, we show that the LF does depend on density, however, we expect the bias caused by the violation of this assumption to be small. In the past, the SWML estimator has been preferred for datasets with smaller sample sizes as it acts to factor out the effect of density perturbations on the LF. 
 
 Broadly, likelihood methods consider the probabilities of observing a galaxy at redshift $z_i$ and magnitude $M_i$ within a magnitude-limited survey. This can be used to construct a likelihood function as follows.

Given the LF, $\Phi(L)$, the probability of observing a galaxy of luminosity $L_i$ at redshift $z_i$ is
\begin{equation}
p_i = \frac{\Phi(L_i)}{\int_{L_\textrm{min}(z_i)}^{L_\textrm{max}(z_i)} \Phi(L)dL} 
,
\end{equation}
giving an overall likelihood for the galaxy sample of 
\begin{equation}
\mathcal{L} = \prod_i p_i  .
\end{equation}

For the SWML method, $\Phi(M)$ is expressed as a set of variables $\Phi_j$ in equally spaced magnitude bins
whose values are then varied using an iterative procedure to maximise the likelihood. This yields
\citep[see][]{1988MNRAS.232..431E}  the \update{ estimator 
\begin{equation} \label{eq:EEP}
    \Phi(M_k) \Delta M = \frac{\sum_{i=1}^N W(M_i - M_k)}{V^{\rm SWML}_k}
\end{equation}
where
\begin{equation}
 V^{\rm SWML} _k={\sum_{i=1}^N \frac{H(M_k -M_{\rm faint}(z_i,k_i)) }{\sum_{j=1}^{N_p} \phi_j  H(M_j - M_\textrm{faint}(z_i,k_i)) \Delta M}},
 \label{eq:VSWML}
\end{equation}
and the binning functions are}
\begin{equation}
  W(x)=\begin{cases}
    1, & \text{if $-\frac{\Delta M}{2} \leq x \leq \frac{\Delta M}{2}$}\\
    0, & \text{otherwise},\\
  \end{cases}
\end{equation}
and
\begin{equation}
  H(x)=\begin{cases}
    0, & \text{if $x \leq -\frac{\Delta M}{2}$}\\
    
    \frac{x}{\Delta L} + \frac{1}{2}, & \text{if $-\frac{\Delta M}{2} \leq x \leq \frac{\Delta M}{2}$}\\

    1, & \text{if $ x \geq \frac{\Delta M}{2}$}.\\
  \end{cases}
\end{equation}
\update{Here $M_{\rm faint}(z_i, k_i)$ represents the faint absolute
magnitude limit of the survey computed at the redshift $z_i$ and with the k-correction $k_i$ of galaxy $i$. 
We adopt a slight modification where the function $H(M_j)$, which
returns the fraction of magnitude bin $j$ that overlaps with
the observed absolute magnitude range,
has been generalized to also include a bright magnitude limit.}

\begin{figure}
    \centering
    \includegraphics[width=\columnwidth]{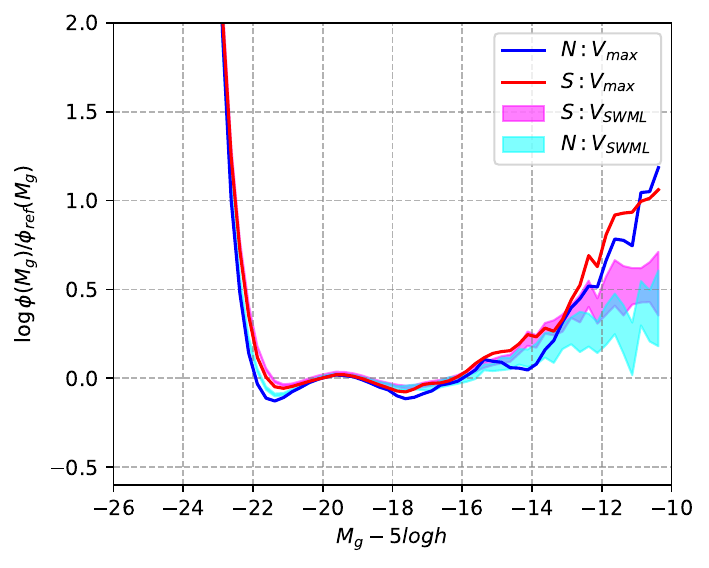}
    \includegraphics[width=\columnwidth]{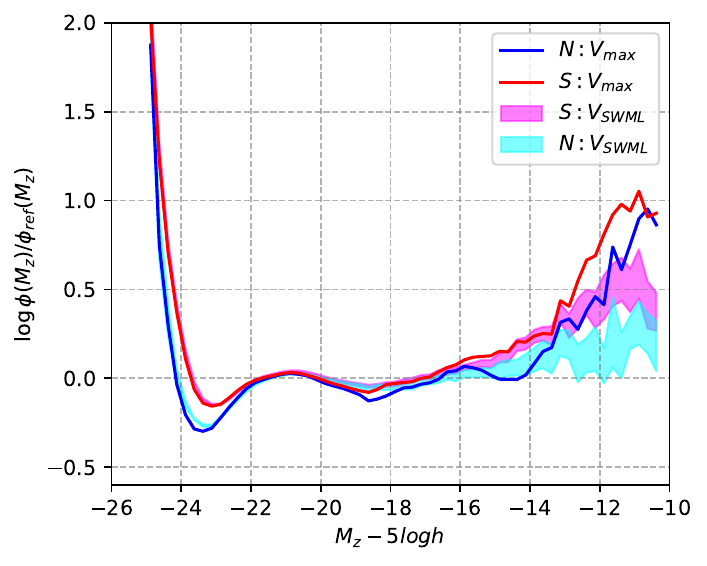}
    \includegraphics[width=\columnwidth]{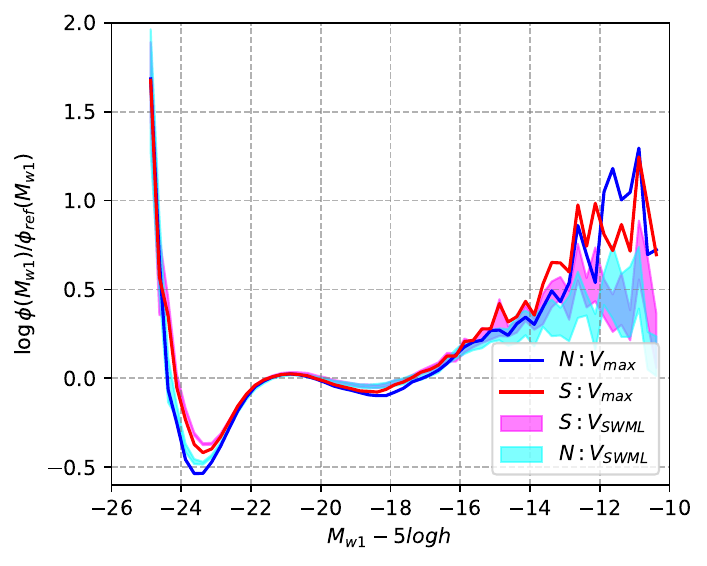}
    \caption{SWML estimates of the $g$, $z$, and $w1$ luminosity functions. All estimates are shown relative to their fiducial Schechter function give in Table~\ref{tab:Sch}. The solid lines show the standard $1/V_\textrm{max}$ estimates (North red and South blue) while the shaded regions show the corresponding SWML estimates and their jackknife errors.
    %\todo{These figures need checking for consistency with new Fig 9 and B1}
    }
    \label{fig:LFgzw1_SWML}
\end{figure}

The density-corrected $V_\textrm{max}$ estimator (hereafter named $V_\textrm{dc,max}$) is based on analysis presented in \cite{2011MNRAS.416..739C}. It is another maximum likelihood iterative method that seeks to correct for density fluctuations in the LF. In this method, an effective or density corrected volume
\begin{equation} \label{eq:effective_volume}
    V_{\textrm{dc,max}, i} = \int_{z_\textrm{min},i}^{z_{\textrm{max},i}} \Delta(z) \frac{dV}{dz} dz = \sum_j \Delta V \Delta_j G(V_j)
\end{equation}
is calculated for each galaxy,
where the sum is over the volume bins, the overdensity parameter is set to $\Delta_j = 1$ in the first iteration, and $G$ is a binning function 
\begin{equation}
  G(V_j)=\begin{cases}
    0, & \text{if $V_j - \Delta V/2 > V_{\textrm{max},i}$}\\
    1, & \text{if $\frac{\min(V_j + \Delta V/2, V_{\textrm{max},i}) - \max(V_j - \Delta V/2, V_{\textrm{min},i})}{\Delta V}$}\\

    0, & \text{if $V_j + \Delta V/2 < V_{\textrm{min},i}$}\\
  \end{cases}.
\end{equation}

For subsequent iterations, the estimate of the overdensity is updated using
\begin{equation}
    \Delta_j = \frac{N_j}{N_{\textrm{exp}, j}},
\end{equation}
where $N_j$ is number of galaxies in volume bin $j$ and
\begin{equation}
    N_{\textrm{exp},j} = \sum_i \frac{H(V_{\textrm{min},i}<V_j<V_{\textrm{max},i}) \Delta V}{\Delta_i V_{\textrm{max}, i}}.
\end{equation}
Here, $N_{\textrm{exp},j}$ is the expectation value for the number of galaxies that we would expect in each volume element if they were uniformly distributed in space given their individual $V_{{\rm min},i}$ and
 $V_{{\rm max},i}$.

From this, the LF for each iteration may be calculated as
\begin{equation}
    \phi = \frac{1}{\Delta M} \sum_{i} \frac{1}{V_\textrm{dc,max}} \Theta(M_i; \Delta M),
\end{equation}
where $\Theta$ is the Heaviside step function. We additionally enforce for each iteration the constraint $\langle \Delta_j \rangle = 1$.

The SWML and $V_{\rm dc, max}$ estimators both remove the effect of density fluctuations by making the assumption that the probability $P(L,z)$ of having a galaxy of luminosity $L$ at redshift $z$ is factorizable as the product of the local density and the shape of the LF. The key difference is that the SWML method starts with the conditional probability distribution $P(L|z)$ and just solves for the shape of the LF, while the $V_{\rm dc, max}$ method starts with the joint $P(L,z)$ and solves for both the shape of the LF and the run of density with redshift.

\update{The SWML method only determines the shape of the LF and not its overall amplitude.
To constrain this we have normalized each LF to have the same luminosity density as the corresponding $1/V_{\rm max}$ LF estimate.}

For each of the three estimators and for North and South separately, 
Fig.~\ref{fig:LF_method} shows the LF estimates 
relative to the corresponding fiducial Schecheter function from
Table~\ref{tab:Sch}. Some small systematic discrepancies exist between the different estimators but most of the features we have previously highlighted remain robust. The bright end of the luminosity functions agree extremely well. The dip below the
fiducial Schechter function around $-17$ persists but the turn up around $-14$ is greatly reduced and consequently likely a result of a density fluctuation.

\update{
For the other bands we compare SWML estimates computed using
\begin{equation}
    \phi(L)dL = \sum_{i=1}^{N} \frac{w_i W(L-L_i)}{V_i^{\rm SWML}},
\end{equation}
with the  corresponding $1/V_{\rm max}$ estimates in 
Fig.~\ref{fig:LFgzw1_SWML}. Here $V_i^{\rm SWML}$ for galaxy $i$
is simply the value given by equation (\ref{eq:VSWML}) for the $r$-band magnitude bin in which galaxy $i$ falls. As in the $r$-band we see good agreement with all features in the $1/V_{\rm max}$
estimates except for at the very faintest magnitudes where in all bands the $1/V_{\rm max}$ has a sharper upturn.
}
\section{Tests Evolution and Comparison to GAMA} \label{appendixLovedaymethods}

%Moved the insertion point of this figure as putting it earlier made it A2 instead of B1
\begin{figure}
    \centering
    \includegraphics[width=\linewidth]{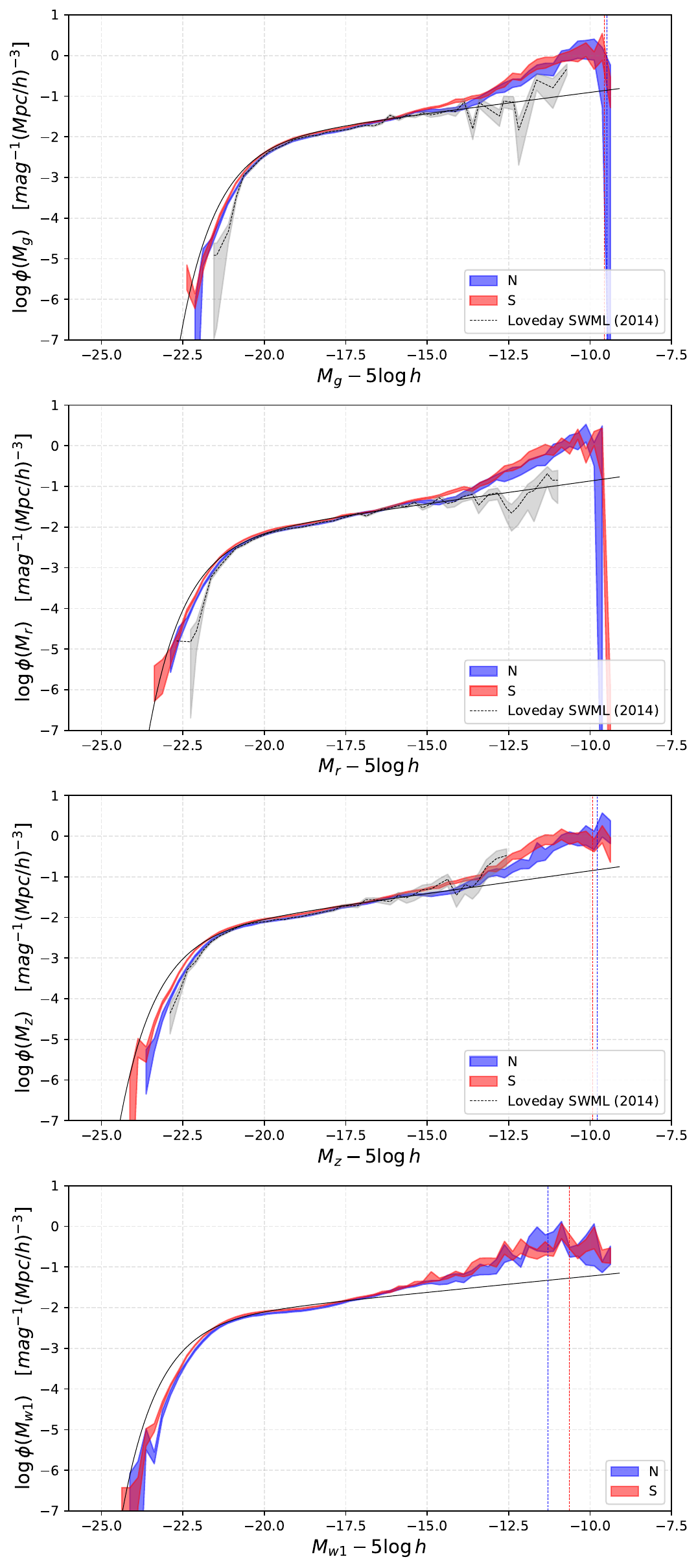}
    \caption{The global $1/V_\textrm{max}$ LF for Y3 data for North and South in the $r$, $g$, $z$ and $w1$ bands in the range $0.002 < z < 0.1$. The absolute magnitudes are calculated with $Q=0$. The width of each LF represents the jackknife error. The dashed lines represent the completeness limits derived from the corresponding bivariate LFs. Schechter functions are the corresponding LFs from Fig.~\ref{fig:global_LF}.}
    \label{fig:global_LF_0p1}
\end{figure}

We would expect our LF estimates to agree well with those of GAMA \citep{Loveday2011}, although with much smaller statistical errors.
In Section~\ref{Results of Global LF estimation} we did not compare our LF estimates with the Loveday LF estimates as \cite{Loveday2011}
make a number of different assumptions.
 Most importantly, the Loveday LFs are defined for a more limited redshift sample (from $ 0.002 < z < 0.1$) with no evolution ($P=0; Q=0$). Fig.~\ref{fig:global_LF_0p1} presents the DESI Y1 BGS LFs in North and South for $0.002<z<0.1$ and $Q=0$. We observe good agreement between the DESI LFs and the Loveday GAMA LFs in the $g$, $r$ and $z$ bands within the GAMA errors over a wide magnitude range. Residual differences at the very bright end are probably due to the different choice of magnitude. GAMA uses Petrosian magnitudes while DESI uses total magnitudes derived
 from model fits. The effect of magnitude choice on the DESI LFs is discussed in Appendix~\ref{appendixFSF}.
 Differences fainter than $-14$ are influenced by local density fluctuations at the very faintest magnitudes some \update{spurious} objects in the DESI catalogue (see Section~\ref{Results of Global LF estimation}).

 In Fig.~\ref{fig:global_LF_0p1} we plot the same fiducial Schechter functions from Table~\ref{tab:Sch} as we did in Fig.~\ref{fig:global_LF}. This enables one to see that there
 are small shifts in the LF estimates that makes those plotted
 here in better agreement with \cite{Loveday2011} than those
 plotted in Fig.~\ref{fig:global_LF}. The reason for this is not due to the different choice evolution parameter $Q$. With these redshift restricted samples, $ 0.002 < z < 0.1$, the LF estimates
 are very insensitive to the choice of $Q$ and we see no discernable difference between estimates with $Q=0$ and $0.78$.
 Instead the small differences between the LF estimates in
 Fig.~\ref{fig:global_LF_0p1} and Fig.~\ref{fig:global_LF}
are due to the failure of the simple evolution model to
fully capture the evolution over the larger redshift range
$ 0.002 < z < 0.6$. We investigate this further below.

\begin{figure}
    \centering
    \includegraphics[width=\linewidth]{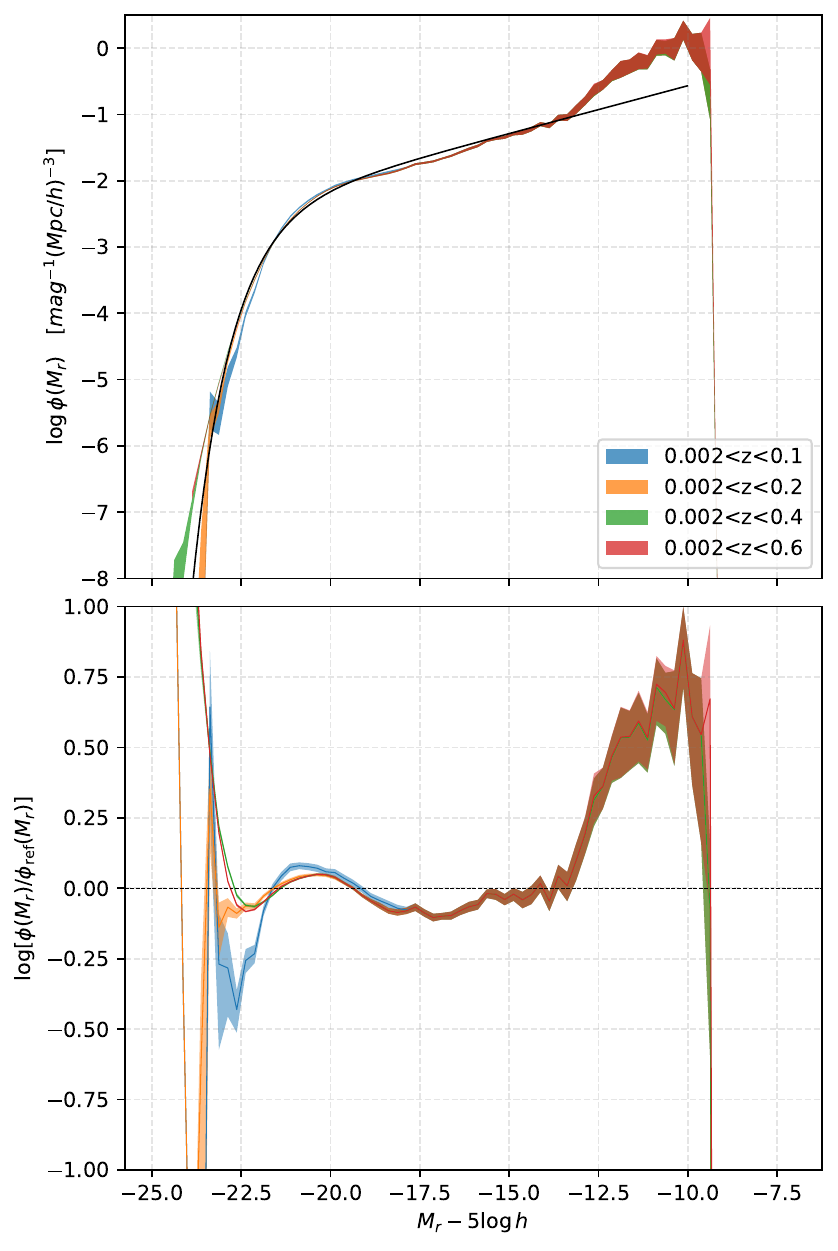}
    \caption{Top: The $1/V_{\rm max}$ total magnitude LFs for different redshift ranges in the South, using $Q=0.78$. Bottom: The ratio plot of each LF to the $r$-band Schechter function in Fig.~\ref{fig:global_LF}.}
    \label{fig:LF_z_comp}
\end{figure}

In Fig.~\ref{fig:LF_z_comp} we plot LF estimates using our standard
$Q=0.78$ evolutionary correction for samples with a range of different upper redshift limits between $z_{\rm max}=0.1$ and
 $z_{\rm max}=0.6$.
The faint parts of the luminosity function necessarily agree perfectly as there are no faint galaxies in the sample above redshift $z=0.1$. The luminosity functions also agree very well
around the knee of the luminosity function. This is to be expected
as the sample we have used to constrain $Q$ is dominated by galaxies around the knee of the luminosity function. Brightward
of the knee we begin to see statistically significant differences
in the estimates. This is particularly clear in the lower panel which shows the ratio of these estimates to our fiducial $r$-band Schechter function from Table~\ref{tab:Sch}.
\update{These differences indicate that a more detailed model of galaxy evolution
is required to model evolution of the LF to the precision warranted
by the small statistical errors of the DESI data. However such modelling would also have to take account of subtle systematics that might exist in the photometric modelling
as the angular size and surface brightness of the sources decrease with redshift.
}

\section{Photometry Difference in North and South} \label{appendixFSF}

\begin{figure}
    \centering
    \includegraphics[width=\columnwidth]{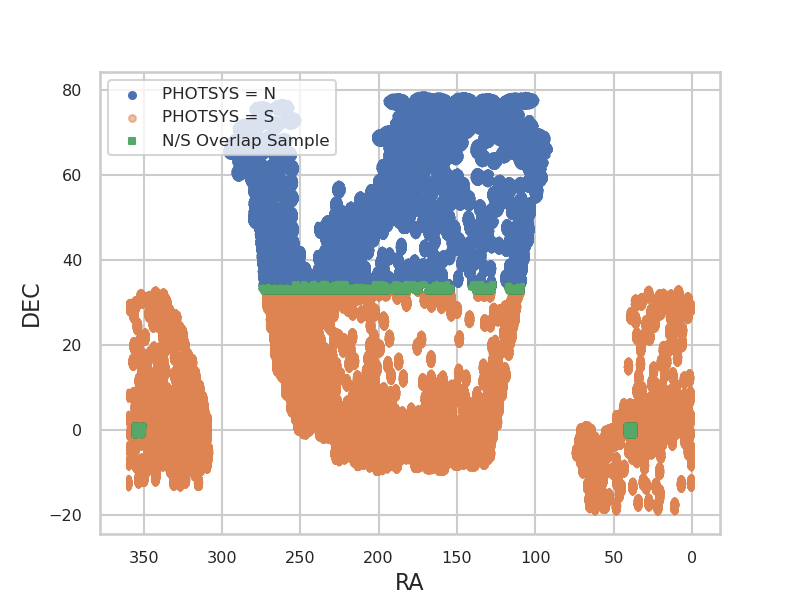}
    \caption[RA-DEC plot showing the North/South overlap.]{A plot showing the North and South regions of the Y1 DESI survey. The green region represents an overlap region - the area where objects received both North and South photometry fluxes (based on BASS/MzLS and DECaLS).}
    \label{fig:FSF_overlap}
\end{figure}

As discussed in Section~\ref{Results of Global LF estimation}, one concern with our LFs is that there is a discrepancy between the North and South $r$-band LFs at the bright end. First, we consider the possibly that there may be an inherent discrepancy in the FSF k-correction fits. To do this, we make use of the overlap region in the DESI survey - that is, the region where there exists both North and South photometry data for the same objects. This region is shown in Figure \ref{fig:FSF_overlap}. 

\begin{figure}
%    \centering
%    \subfloat{
    \includegraphics[width=1.0\columnwidth]{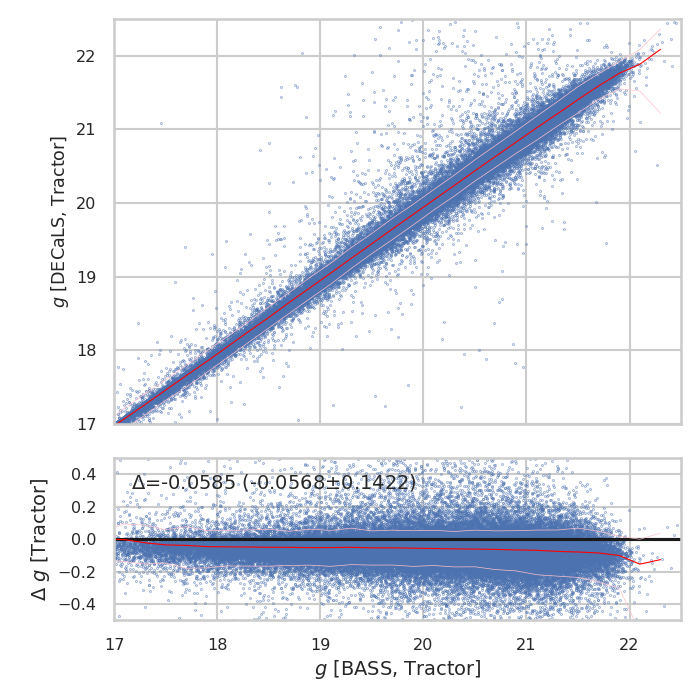}
    %}
%    \subfloat{
    \includegraphics[width=1.02\columnwidth]{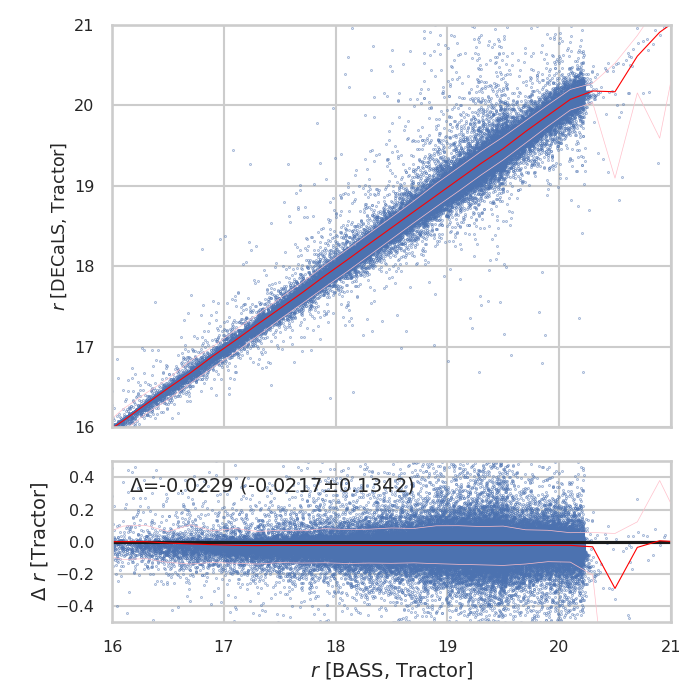}
    %}
    \caption[Comparison of BASS and DECaLS apparent magnitudes.]{Comparison of BASS and DECaLS apparent magnitudes. The upper pair of panels are for the $g$-band and the lower pair for the $r$-band apparent magnitudes. In each case the the lower sub-panel shows the difference (DeCALS-BASS) 
     versus the BASS magnitude. In all panels the red lines shows the median and the light lines the 5th and 95th percentiles of the distributions. The values in the lower plots are the median across the whole sample and, in brackets, the mean and standard deviation of the difference.}
    \label{fig:rg-BASS_DECALS}
\end{figure}

In Fig.~\ref{fig:rg-BASS_DECALS}, we compare the apparent magnitudes ($g$ and $r$) in BASS and DECaLS. As the effective photometric pass bands are not the same for BASS and DeCALS we expect some systematic differences between the two magnitudes.
This can clearly be seen in the $g$-band where there is a fairly constant offset in the median of 0.059 magnitudes. The $r$-band filters are much more similar and here the corresponding median offset is 0.023 magnitudes. Because of these differences it is 
better to compare their absolute magnitudes using the k-corrections
described in Section~\ref{k-corrections and e-corrections}
which take the passbands into account and convert to a consistent
SDSS $z_\textrm{ref}=0.1$ rest-frame passband.

For blue galaxies the two magnitudes agree well over the full absolute magnitude range with a median magnitude offset of only 0.01.  However for the red galaxies we only see good agreement between the two magnitudes for galaxies fainter than $^{0.1}M_\textrm{r}-5 \log h=-21$. At bright magnitudes we see an increasing
offset that grows to 0.1 magnitudes
at $^{0.1}M_\textrm{r}-5 \log h=-23$ magnitudes.
We  note that there exist a sufficient number of blue galaxies at
these bright magnitudes to demonstrate that this does not hold for blue galaxies.

\begin{figure}
    %\centering
    %\subfloat{{
     \includegraphics[width=1.0\columnwidth]{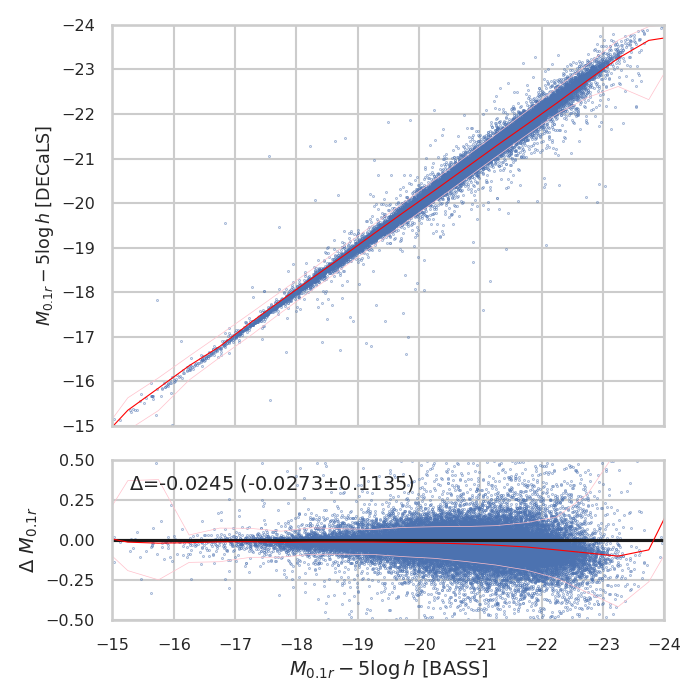} %}}
    %\subfloat {
    \includegraphics[width=1.0\columnwidth]{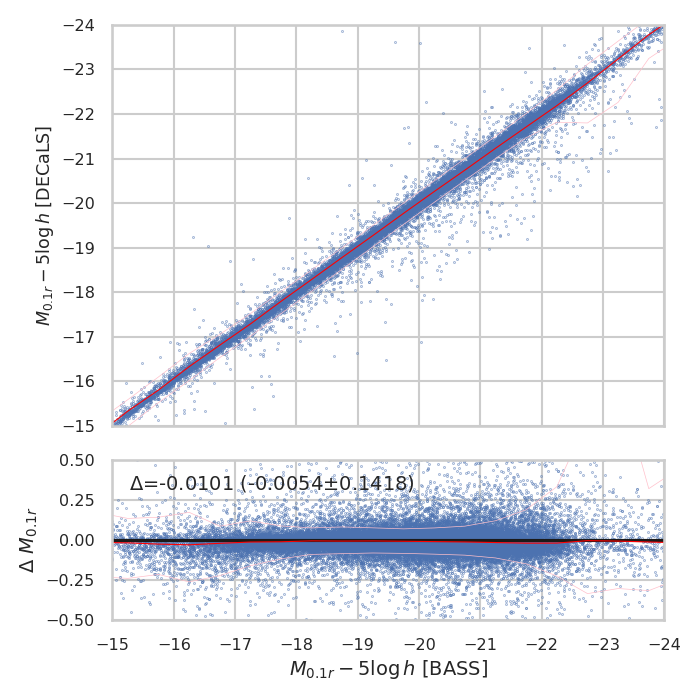} %}}
    \caption[Comparison of BASS and DECaLS $r$-band absolute magnitudes for red and blue galaxies.]{
    Comparison of BASS and DECaLS $r$-band absolute magnitudes 
    k-corrected to the SDSS $z_\textrm{ref}=0.1$ passband
    for red and blue galaxies.
    The upper pair of panels is for the red galaxies ($^{0.1}(g-r) > 0.75$) and the lower pair for blue galaxies ($^{0.1}(g-r) < 0.75$). The lower panel in each pair shows the residual (DECaLS-BASS) versus the BASS magnitude.  In all panels the red lines shows the median and the light lines the 5th and 95th percentiles of the distributions. The values in the lower plots are the median across the whole sample and, in brackets, the mean and standard deviation of the difference.}
    \label{fig:Mr_BASS_DECALS}
\end{figure}

We have seen in Fig~\ref{fig:colour_LF} the discrepancy between the North and South luminosity functions is confined to the bright end of the
red galaxy luminosity function.  The offset that we have found above is
in the sense that could help explain this discrepancy.
To investigate this quantitatively we have assigned new magnitudes to the red galaxies in the North by shifting them by the magnitude-dependent median difference found in the overlap region such that
\begin{equation}    
M^\textrm{BASS}_{r,i, \textrm{new}} = M^\textrm{BASS}_{r,i,\textrm{old}} + 
\left\langle 
M^\textrm{DECaLS}_{r} - M^\textrm{BASS}_{r}
\right\rangle_{M^\textrm{BASS}_{r,i,\textrm{old}}}
\end{equation}

\begin{figure}
    \centering
    \subfloat{{\includegraphics[width=\columnwidth]{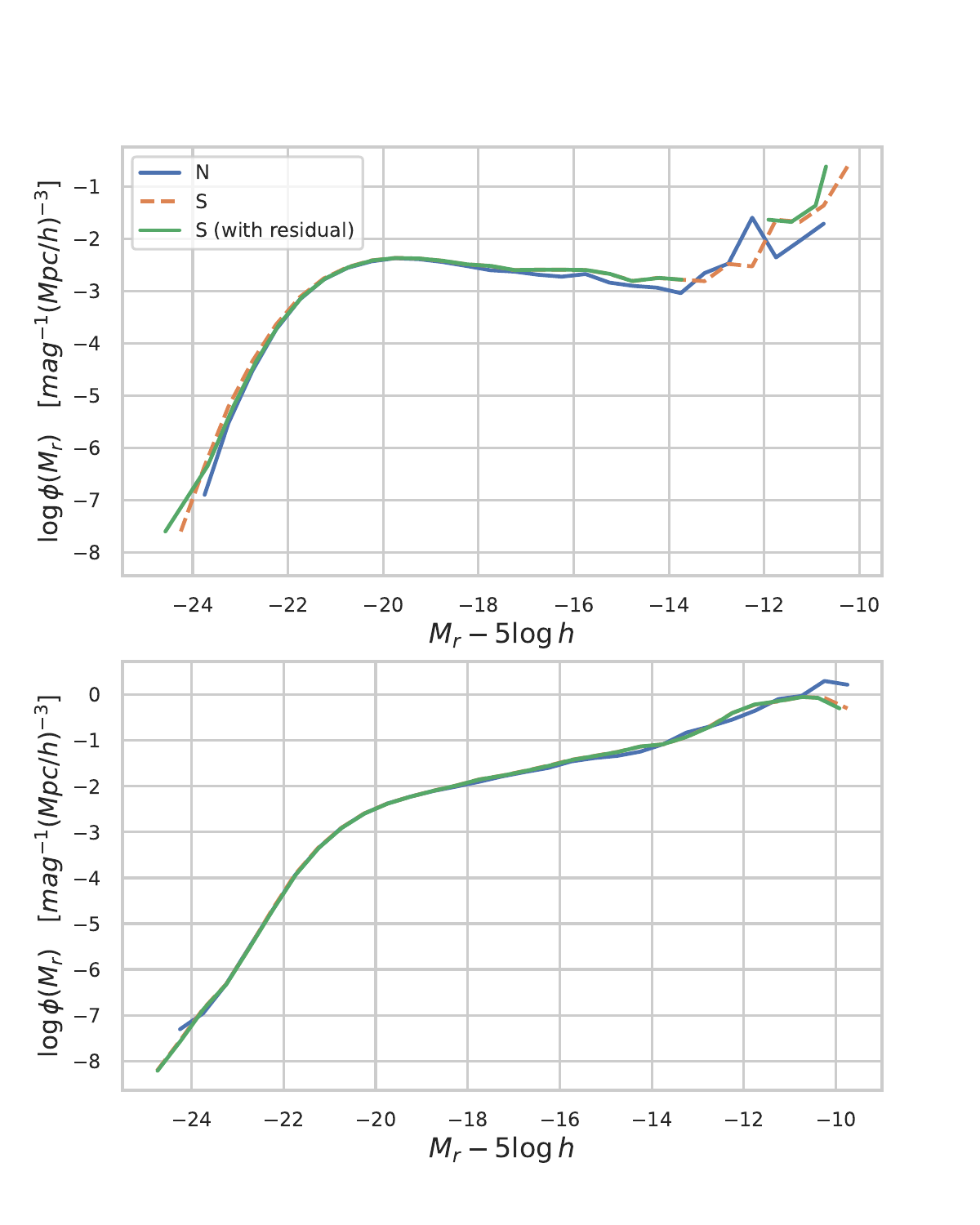}}}
    
    \caption[LFs for red and blue galaxies with BASS/DECaLS offset.]{Top: a plot of the North and South $r$-band LFs for red galaxies. Here each galaxy in the South has its $^{0.1}M_r - 5\log h$ changed by the residual of the BASS/DECaLS $^{0.1}M_r - 5\log h$ for that magnitude bin. This yields the South LF ‘with residual' offset. Bottom: The same LFs but for the blue galaxies.}
    \label{fig:perturbed_LF}
\end{figure}

The effect of this adjustment on the $r$-band luminosity function
can be seen Fig.~\ref{fig:perturbed_LF}. The adjustment makes a notable difference to the LF at bright magnitudes, bringing the North LF far closer to that of the South. While we do not expect this adjustment to fully correct the LF since it only takes account of the median offset and not the cause of the scatter it nevertheless indicates that the difference in the bright ends of the $r$-band LFs is consistent with brighter red galaxies being systematically assigned different magnitudes dependent on whether the North or South photometry is used. It does not explain why this is the case or why we see similar offsets in other bands and so we have investigated other properties of the bright red galaxies.

\begin{figure}
    \centering
    \includegraphics[width=\linewidth]{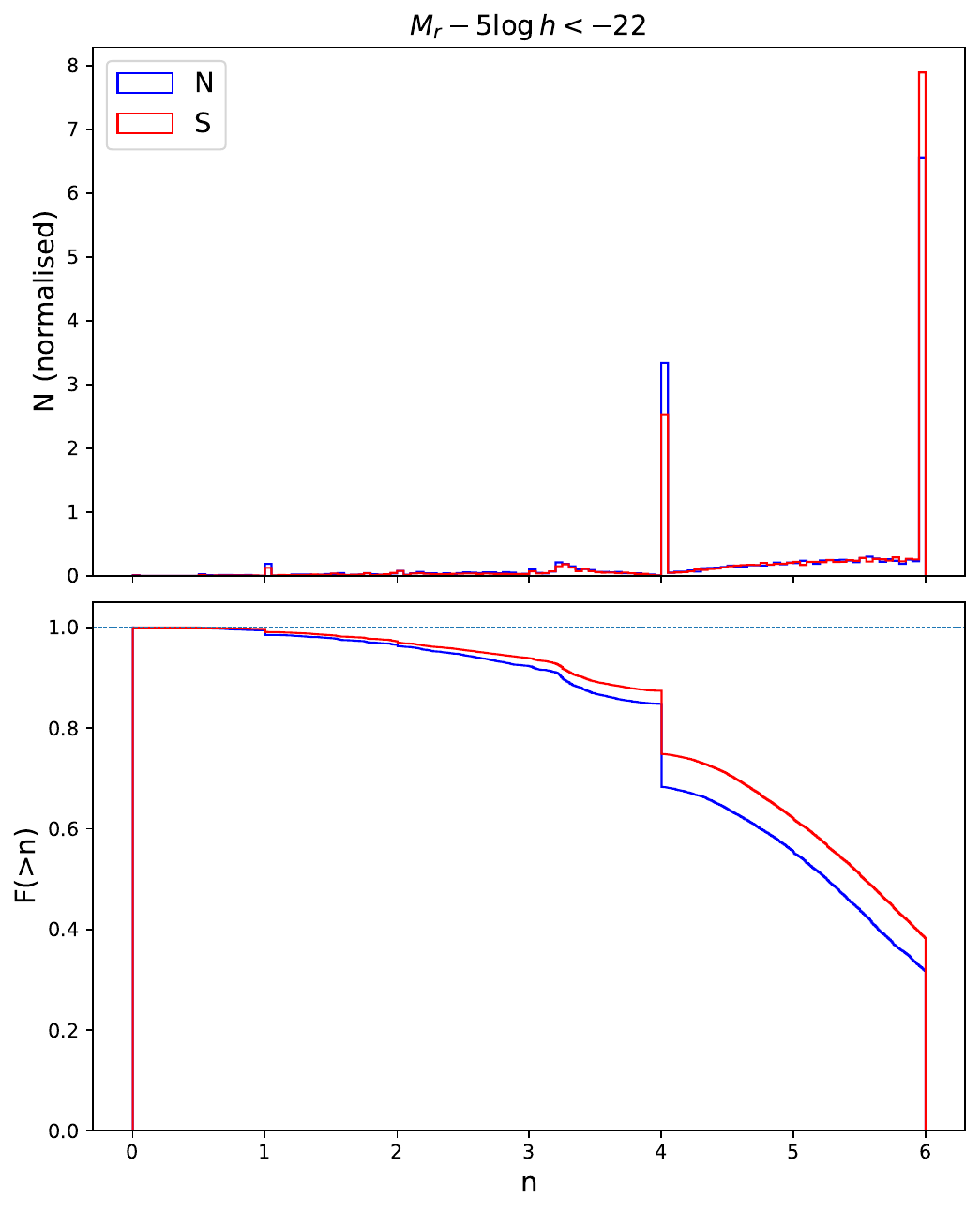}
    \caption{Top: the normalised distribution of the Sérsic index. Bottom: The cumulative distribution plots of the Sérsic index, $n$, for galaxies brighter than $^{0.1}M_r - 5\log h < -22$ in the North and South.}
    \label{fig:cdf_Sérsic}
\end{figure}

Bright red galaxies are typically early-type galaxies with 
light profiles that are of the de Vaucouleurs form (Sérsic profile with index $n=4$) or have even higher Sérsic indices. It is therefore interesting to see if the distribution of Sérsic index for bright galaxies is consistent in the North and South regions.
Fig.~\ref{fig:cdf_Sérsic} compares normalized histograms of the
distribution of Sérsic index and the corresponding cumulative fractions, $F(<n)$, for galaxies brighter than  $^{0.1}M_r - 5\log h < -22$, which is the magnitude range where the North and South luminosity functions disagree. The first thing we see is that the
spikes at $n=1$ (exponential profile) and $n=4$ (de Vaucouleurs profile) are smaller in the South than the North. Our understanding of this is that the TRACTOR photometric pipeline only adopts a Sérsic profile if the $\Delta \chi^2$ between its fit and the fit of the simpler exponential/de Vaucouleurs profile is above a threshold that warrants the inclusion of the extra parameter. As the South/DECaLS data is deeper than the North/BASS data the typical signal-to-noise is greater and this threshold will be passed more often in the South. This is not a concern as presumably the value of $n$ found should still be close to the $n=1$, or $n=4$ values. However we also see a difference in the distributions at values of $n>4$.

The maximum value of the Sérsic index that is considered is $n=6$. In the South we see there are more galaxies fit with this maximum allowed value than in the North. Moreover this difference
is not just restricted to distortions in the distribution close to $n=6$ as the cumulative distribution shows the difference persists to $n<4$. One possibility is that in the deeper South data 
provides more information in the low surface 
brightness wings of the most extended galaxies 
and this favours larger $n$ than is warranted
in the noisier North data.
One way to investigate this would
be to compare fixed aperture magnitudes rather model fits but these are not available. As an alternative, we have computed
Petrosian magnitudes from the model fits \citep{2001AJ....121.2358B}. Petrosian magnitudes are aperture magnitudes defined within an aperture
that scales with the scale of the light profile of the galaxy
and are independent of the light distribution beyond this radius.

\begin{figure}
    \centering
    \includegraphics[width=\linewidth]{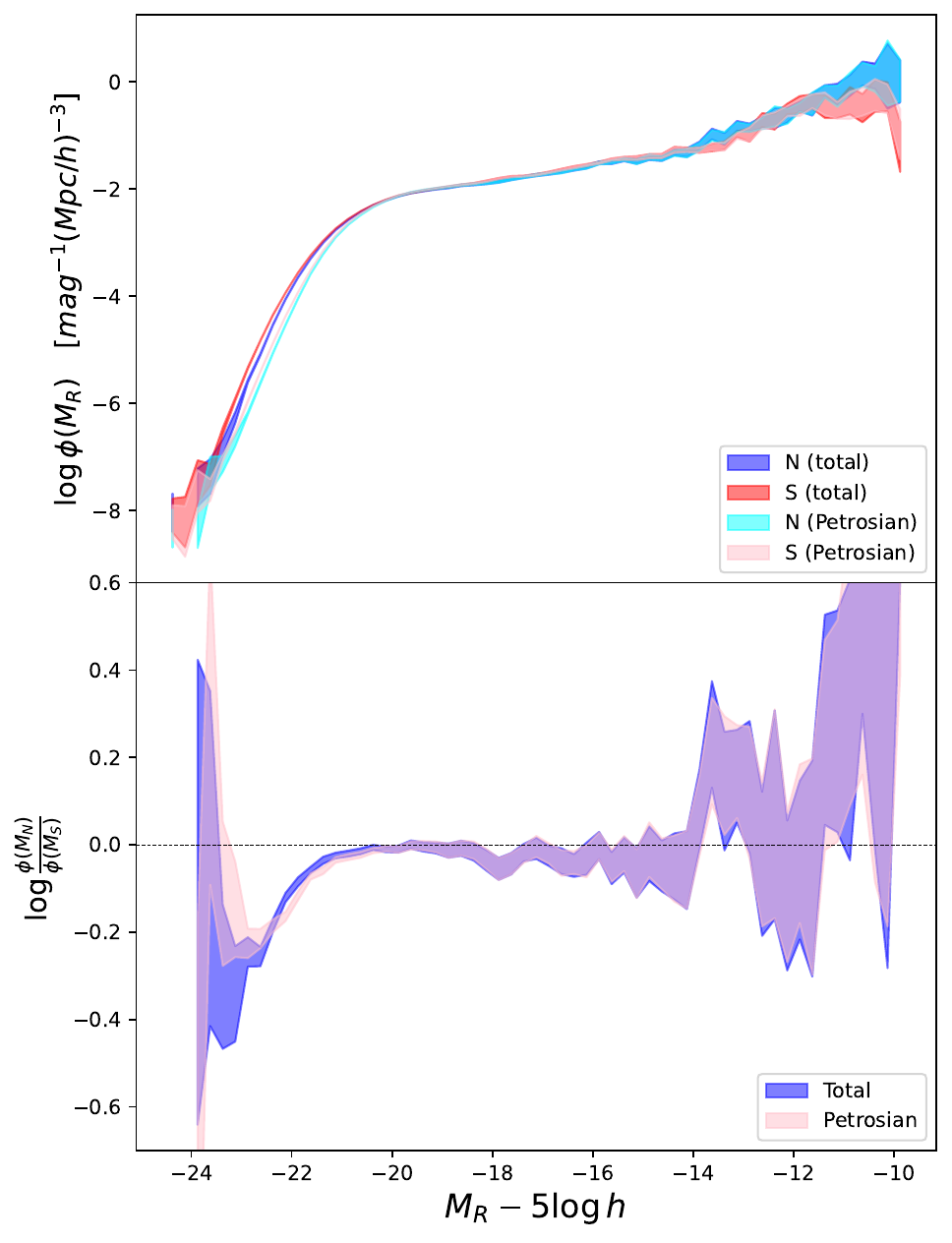}
    \caption{Top: The total magnitude $r$-band LFs plotted alongside the Petrosian magnitude LFs 
    for samples selected within the range $0.002 < z < 0.6$. All absolute magnitudes are calculated with $Q=0.78$. Bottom: The ratio the North and South LFs. In blue is shown the ration for the
    total magnitude LFs and in pink those for the Petrosian magnitudes.
     The width of each band represents the jackknife error estimate.}
    \label{fig:Petrosian_LF}
\end{figure}

We do not have the light profiles to compute true Petrosian magnitudes and so instead we have computed Petrosian magnitudes from the \update{2D elliptical Sérsic fits. The radial coordinate of these model profiles, denoted $r$, is the semi-major axis of the ellipse.}
For a Sérsic profile, the ratio of the Petrosian flux to the total flux is given by
\begin{equation}
\frac{F_\textrm{p}}{F_\textrm{tot}}=\frac{\int_0^{2r_\textrm{p}} \exp\left(- r ^{1/n} \right) r \, dr}{\int_0^{\infty} \exp\left(- r ^{1/n} \right)   r' \, dr'},
\end{equation}
where the Petrosian radius, $r_\textrm{p}$, is the root of equation
\begin{equation}
\eta(r)=\frac{r^2 \exp\left(- r^{1/n} \right)}{2\int_0^r \exp\left(- r'^{1/n} \right)   r' \, dr'} =0.2 .
\end{equation}

As with true Petrosian magnitudes, we believe these magnitudes should be less sensitive to the outer profile of 
the galaxy.
 As such, we hypothesise that North and South LFs based on Petrosian absolute magnitudes will converge at the bright-end. 

To estimate the LF using these Petrosian magnitudes 
we must define a new sample that is complete with respect to Petrosian magnitude limits.  As the 
Petrosian magnitudes are all fainter than the corresponding total magnitude some objects will be fainter than 
 the BGS Bright target selection faint magnitude limit. Hence, we discard objects by applying $r_{\rm petro} < 19.5$ for the South, and $r_{\rm petro} < 19.54$ for the North. We additionally recompute all required values (such as the $V_{\rm max}$ values) as detailed above.
 
 Fig.~\ref{fig:Petrosian_LF} presents these Petrosian-magnitude LFs alongside those that use the standard TRACTOR total magnitudes. Faintward of magnitude -20
 the two types of luminosity function agree extremely well. This is to be expected as we have seen that this part of the LF is dominated by blue galaxies which are typically late-type galaxies with exponential light profiles. As such light profiles have very little flux outside the Petrosian radius the LFs will necessarily agree very well. In contrast, at the bright end the Petrosian LFs are shifted faintward consistent with this part of the luminosity function being dominated by red late-type galaxies with extended light profiles for which there is significant flux beyond the Petrosian radius.  Also of interest is that the Petrosian LFs
 are in \update{moderately} better agreement in North and South but as can be seen in the lower panel of
 Fig.~\ref{fig:Petrosian_LF} this change 
does not fully remove the discrepancy between the 
bright ends of the North and South LFs.
Nevertheless this analysis does suggest that the biggest systematic uncertainty in the bright end of the galaxy luminosity function is related to quantifying the low surface brightness outskirts of
bright red early-type galaxies. We would expect the difference in North and South LFs to decrease further if true Petrosian magnitudes were used rather than 
our model Petrosian magnitudes which are still somewhat influenced by the fits to the outer profile.

Whilst this analysis could be taken to indicate that the disparity in the LFs at the bright end is due to flux being missed in
 the North data due to erroneous Sérsic profile fitting, we caution against this interpretation. We note the alternative possibility that the very high Sérsic indices found in the South may be an indication that the formally smaller photometric errors make the profile fitting more susceptible to other systematic errors such as residual errors in the sky subtraction. Without further 
 investigation, we are unable to say which LF is the correct one. Instead, we interpret the difference as the level of systematic uncertainty that remains in the estimate of the bright end of the galaxy luminosity function.

\section{Imaging Visual Inspection} \label{tableVIimaging}

\update{Our LF estimates reach extremely faint magnitudes across all bands. Whilst we have outlined the methodology for estimating the completeness of the $g$, $z$, and $w1$ bands using the bivariate LFs, we caution that this is a theoretical completeness limit that does not consider other factors that may introduce error in the faint-end of the LF (including for the $r$-band LF). In order to better understand the completeness of our data, we conduct a visual inspection programme of the DESI Y3 galaxies using the Legacy Survey Sky Browser. In particular, we are interested as to whether the galaxies at very faint $r$-band absolute magnitudes are real galaxies with accurate fluxes. We also used this investigation to inform the choice of the $r-w1<2.25$ colour cut we applied to remove problematic galaxies from the
the to the sample used to estimate the $w1$ LF.

The Legacy Survey Sky Browser is a viewing tool that allows us to visually inspect the imaging from various surveys across the whole sky. This viewer usefully contains the Legacy Survey DR9 images in the $ugriz$ bands as well as the unWISE W1/W2 NEO7 images, allowing for quick visual comparisons of selected galaxies. In addition, these images can be overlayed with DESI information, including the DESI Bright-time targets and the DESI footprint and fibres. We constructed a script to extract two sets of Sky Browser DR9 images each comprising of 20 random objects within each magnitude bin.  
The script compiles the images and overlays details such as the RA, DEC, $z$, and TARGETID. From this, visual inspection may quickly be conducted for each magnitude bin. 

\begin{table}
\caption{\small{Table showing the visual classification of two independent sets of randomly selected objects (A \& B) in bins of $r$-band magnitude. The two sets were classified by different members of the team.
The columns $M_r^{\rm faint}$ and $M_r^{\rm bright}$ define the $M_r - 5 \log h$ range. Columns  3 \& 4 (5 \& 6) indicate how many galaxies in the sample A (B) appear to be well modelled galaxies (Good) and how many appear problematic (Bad).}}
\begin{tabular}{llllll}
$M^\textrm{faint}_r$ & $M^\textrm{bright}_r$ & Good (A) & Bad  (A) & Good (B)& Bad (B)\\ \hline
%       & -10.0  & 12  & 29 \\
%-10.00 & -10.25 & 12  & 8  \\
-10.25 & -10.5  & 7   & 13 & 6 & 14\\
-10.5  & -10.75 & 8   & 12 & 3  & 17 \\
-10.75 & -11.0  & 12  & 13 &  11 & 9\\
-11.0  & -11.25 & 13  & 7  &  10 & 10\\
-11.25 & -11.5  & 15  & 5 &  8 & 12\\

-11.5  & -11.75 & 9   & 11 &  5 & 15 \\
-11.75 & -12.0  & 15  & 5  &  11  & 9\\
-12.0  & -12.25 & 10  & 10 &  10 &10 \\
-12.25 & -12.5  & 13  & 7  &  8 & 12\\
-12.5  & -12.75 & 18  & 2 &  8 & 12 \\
-12.75 & -13.0  & 12  & 8  & 10 &10 \\
-13    & -13.25 & 12  & 8 &  11 & 9\\

-13.25 & -13.5  & 15  & 5 &  12& 8\\
-13.5  & -13.75 & 14  & 6  & 15 & 5\\
-13.75 & -14.0  & 15  & 5  & 17 &3\\
-14.0  & -14.25 & 12  & 8  & 17 &3\\
-14.25 & -14.5  & 16  & 4 &  15&  5\\
-14.5  & -14.75 & 15  & 5 &  17 & 3\\
-14.75 & -15.0  & 14  & 6 &  16 & 4\\

-15.0  & -15.25 & 17  & 3 &  18 & 2\\
-15.25 & -15.5  & 14  & 6  &  20 & 0\\
-15.5  & -15.75 & 18  & 2  &  17 & 3\\
-15.75 & -16.0  & 18  & 2 &  18 & 2\\
-16.0  & -16.25 & 18  & 2  & 15  & 5\\
-16.25 & -16.5  & 20  & 0  &  19 & 1\\
-16.5  & -16.75 & 19  & 1  &  20 & 0\\

-16.75 & -17.0  & 20  & 0  &  20 & 0\\
-17.0  & -17.25 & 20  & 0  &  19 & 1\\
-17.25 & -17.5  & 20  & 0  &  20 & 0\\
-17.5  & -17.75 & 19  & 1  &  20 & 0\\
-17.75 & -18.0  & 20  & 0  &  20 & 0\\
-18.0  & -18.25 & 19  & 1 &  20& 0\\
-18.25 & -18.5  & 20  & 0 &  19 & 1\\ \hline
\end{tabular}
\label{tab:imaging}\end{table}

We then define two categories: “good” and “bad”. Good galaxies are characterised by being visible, likely galaxies. Typically they will be the dominant object targeted by the fibre, and although there may be other similar objects or features in the image that could conceivably affect the flux extraction, we are confident that the object targeted is a galaxy. Bad galaxies on the other hand are objects that are unlikely to be galaxies, or are galaxies that are obviously incorrectly modelled. The vast majority of bad galaxies are shredded components of much larger galaxies. 
This issue and ways of correcting the galaxy photometry are investigated further in Manwadkar
et al (in prep).

\begin{figure}
    \centering
    \includegraphics[width=\columnwidth]{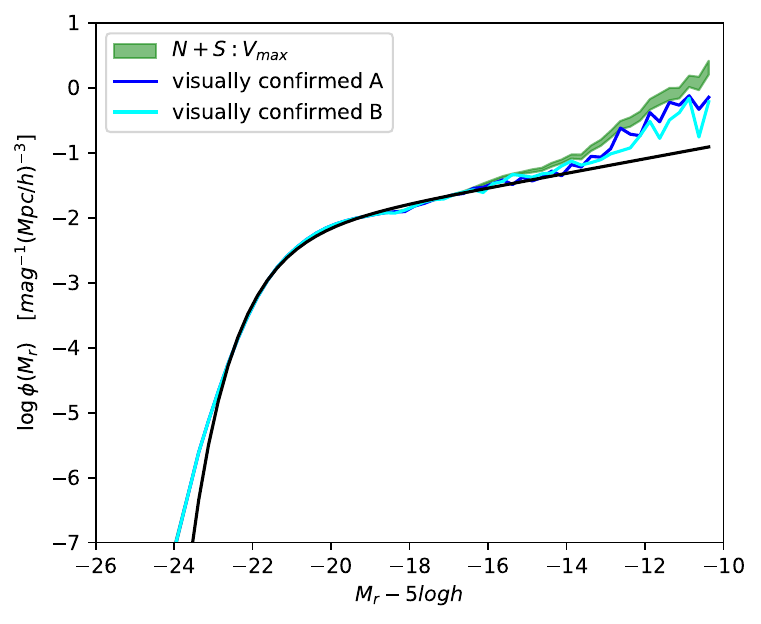}
    \includegraphics[width=\columnwidth]{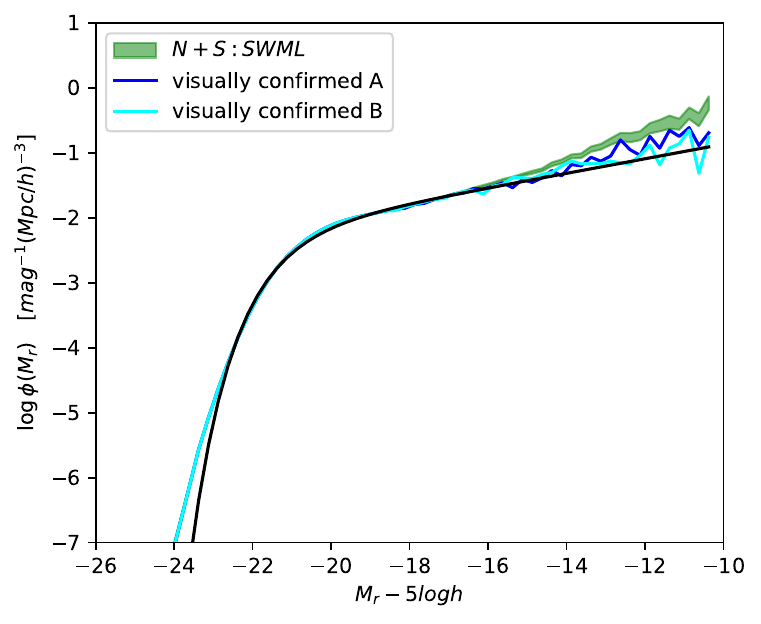}
    \caption{The area weighted average of the North and South $1/V_\textrm{max}$ (top) and SWML (bottom) LFs are shown in green. The width of the band  shows the jackknife $\pm 1\sigma$ error range.
  The effect on this LF of weighting by the visually confirmed fraction 
  in sample~A and~B are shown by the lines as indicated in the legend. }
    \label{fig:VIweighted}
\end{figure}

Table~\ref{tab:imaging} shows our classification results for both sample~A and~B which were classified by different members of the team.
We investigate the impact of these problem targets on the $r$-band LF, by weighting by the fraction of galaxies classified as good in each magnitude bin. The result of this is shown in Fig.~{\ref{fig:VIweighted} which shows both $1/V_\textrm{max}$ and SWML LF estimates. 
The confirmed fraction agrees well between the two samples at absolute magnitudes brighter than $-14$. At fainter magnitudes the confirmed fraction is slightly lower in sample B than sample A. This difference is as much down to the subjective nature of the visual classification as it is to the small sample size.  Nevertheless it is clear by both estimates that shredded galaxies are contributing significantly to the upturn we have observed in the faint end of the luminosity function. In the case of the SWML LF estimate, which also reduces the strength of the faint end upturn by correcting for local density fluctuations, the estimate corrected using sample B shows no upturn. Hence we conclude that within the accuracy of our estimates the
faint upturn in the $1/V_\textrm{max}$ LF is caused by a combination of local overdensity combined with galaxies that are photometrically shredded to produce spurious faint targets. 
}

}

\vfill

\bibliographystyle{mnras} 
\bibliography{refs}

\clearpage
\update{
{\itshape
% List of institutions
\noindent
$^{1}$Institute for Computational Cosmology, Department of Physics, Durham University, South Road, Durham DH1 3LE, UK\\
$^2$Department of Astrophysical Sciences, Princeton University, 4 Ivy Lane, Princeton, NJ 08544, USA\\
$^{3}$Centre for Extragalactic Astronomy, Department of Physics, Durham University, South Road, Durham DH1 3LE, UK\\
$^{4}$Department of Physics and Astronomy, Siena College, 515 Loudon Road, Loudonville, NY
12211, USA \\
$^5$Lawrence Berkeley National Laboratory, 1 Cyclotron Road, Berkeley, CA 94720, USA \\
$^{6}$Department of Physics, Boston University, 590 Commonwealth Avenue, Boston, MA 02215 USA \\
$^{7}$Dipartimento di Fisica ``Aldo Pontremoli'', Universit\`a degli Studi di Milano, Via Celoria 16, I-20133 Milano, Italy \\
$^{8}$INAF-Osservatorio Astronomico di Brera, Via Brera 28, 20122 Milano, Italy \\
$^{9}$Department of Physics \& Astronomy, University College London, Gower Street, London, WC1E 6BT, UK \\
$^{10}$Institut d'Estudis Espacials de Catalunya (IEEC), c/ Esteve Terradas 1, Edifici RDIT, Campus PMT-UPC, 08860 Castelldefels, Spain \\
$^{11}$Institute of Space Sciences, ICE-CSIC, Campus UAB, Carrer de Can Magrans s/n, 08913 Bellaterra, Barcelona, Spain \\
$^{12}$Instituto de F\'{\i}sica, Universidad Nacional Aut\'{o}noma de M\'{e}xico, Circuito de la Investigaci\'{o}n Cient\'{\i}fica, Ciudad Universitaria, Cd. de M\'{e}xico  C.~P.~04510,  M\'{e}xico \\
$^{13}$NSF NOIRLab, 950 N. Cherry Ave., Tucson, AZ 85719, USA \\
$^{14}$Department of Astronomy \& Astrophysics, University of Toronto, Toronto, ON M5S 3H4, Canada \\
$^{15}$Department of Physics \& Astronomy and Pittsburgh Particle Physics, Astrophysics, and Cosmology Center (PITT PACC), \\University of Pittsburgh, 3941 O'Hara Street, Pittsburgh, PA 15260, USA \\
$^{16}$University of California, Berkeley, 110 Sproul Hall \#5800 Berkeley, CA 94720, USA \\
$^{17}$Institut de F\'{i}sica dâ Altes Energies (IFAE), The Barcelona Institute of Science and Technology, Edifici Cn, Campus UAB, 08193, Bellaterra (Barcelona), Spain \\
$^{18}$Departamento de F\'isica, Universidad de los Andes, Cra. 1 No. 18A-10, Edificio Ip, CP 111711, Bogot\'a, Colombia \\
$^{19}$Observatorio Astron\'omico, Universidad de los Andes, Cra. 1 No. 18A-10, Edificio H, CP 111711 Bogot\'a, Colombia \\
$^{20}$Institute of Cosmology and Gravitation, University of Portsmouth, Dennis Sciama Building, Portsmouth, PO1 3FX, UK \\
$^{21}$University of Virginia, Department of Astronomy, Charlottesville, VA 22904, USA \\
$^{22}$Fermi National Accelerator Laboratory, PO Box 500, Batavia, IL 60510, USA \\
$^{23}$Institut d'Astrophysique de Paris. 98 bis boulevard Arago. 75014 Paris, France \\
$^{24}$IRFU, CEA, Universit\'{e} Paris-Saclay, F-91191 Gif-sur-Yvette, France \\
$^{25}$Center for Cosmology and AstroParticle Physics, The Ohio State University, 191 West Woodruff Avenue, Columbus, OH 43210, USA \\
$^{26}$Department of Physics, The Ohio State University, 191 West Woodruff Avenue, Columbus, OH 43210, USA \\
$^{27}$The Ohio State University, Columbus, 43210 OH, USA \\
$^{28}$Department of Physics, The University of Texas at Dallas, 800 W. Campbell Rd., Richardson, TX 75080, USA \\
$^{29}$Department of Physics, Southern Methodist University, 3215 Daniel Avenue, Dallas, TX 75275, USA \\
$^{30}$Institute for Astronomy, University of Edinburgh, Royal Observatory, Blackford Hill, Edinburgh EH9 3HJ, UK \\
$^{31}$Institute of Astronomy, University of Cambridge, Madingley Road, Cambridge CB3 0HA, UK \\
$^{32}$Sorbonne Universit\'{e}, CNRS/IN2P3, Laboratoire de Physique Nucl\'{e}aire et de Hautes Energies (LPNHE), FR-75005 Paris, France \\
$^{33}$Departament de F\'{i}sica, Serra H\'{u}nter, Universitat Aut\`{o}noma de Barcelona, 08193 Bellaterra (Barcelona), Spain \\
$^{34}$Instituci\'{o} Catalana de Recerca i Estudis Avan\c{c}ats, Passeig de Llu\'{\i}s Companys, 23, 08010 Barcelona, Spain\\
$^{35}$Department of Physics and Astronomy, University of Waterloo, 200 University Ave W, Waterloo, ON N2L 3G1, Canada \\
$^{36}$Perimeter Institute for Theoretical Physics, 31 Caroline St. North, Waterloo, ON N2L 2Y5, Canada \\
$^{37}$Waterloo Centre for Astrophysics, University of Waterloo, 200 University Ave W, Waterloo, ON N2L 3G1, Canada \\
$^{38}$Space Sciences Laboratory, University of California, Berkeley, 7 Gauss Way, Berkeley, CA  94720, USA \\
$^{39}$Instituto de Astrof\'{i}sica de Andaluc\'{i}a (CSIC), Glorieta de la Astronom\'{i}a, s/n, E-18008 Granada, Spain \\
$^{40}$Department of Astronomy, The Ohio State University, 4055 McPherson Laboratory, 140 W 18th Avenue, Columbus, OH 43210, USA \\
$^{41}$Department of Physics and Astronomy, Sejong University, 209 Neungdong-ro, Gwangjin-gu, Seoul 05006, Republic of Korea \\
$^{42}$CIEMAT, Avenida Complutense 40, E-28040 Madrid, Spain \\
$^{43}$Department of Physics, University of Michigan, 450 Church Street, Ann Arbor, MI 48109, USA \\
$^{44}$University of Michigan, 500 S. State Street, Ann Arbor, MI 48109, USA \\
$^{45}$Department of Physics \& Astronomy, Ohio University, 139 University Terrace, Athens, OH 45701, USA \\
$^{46}$Kavli Institute for Particle Astrophysics and Cosmology, Stanford University, Menlo Park, CA 94305, USA \\
$^{47}$Physics Department, Stanford University, Stanford, CA 93405, USA \\
$^{48}$SLAC National Accelerator Laboratory, 2575 Sand Hill Road, Menlo Park, CA 94025, USA \\
$^{49}$National Astronomical Observatories, Chinese Academy of Sciences, A20 Datun Road, Chaoyang District, Beijing, 100101, P.~R.~China 
}}

\label{lastpage}
\end{document}